\documentclass{jfm}
\usepackage{graphicx}
\usepackage{epsfig}

\usepackage{amsmath}
\usepackage{float}
\usepackage{siunitx}
\usepackage{breqn}
\usepackage[position=top]{subfig}
\usepackage{mathrsfs}
\usepackage{caption}

\usepackage{xspace}

\def\fixedlabel#1#2{%
  \@bsphack%
  \protected@write\@auxout{}%
         {\string\newlabel{#1}{{#2}{\thepage}}}%
  \@esphack}
 
\def\Re{\text{Re}}
\def\mi{\mathrm{i}} 
\def\R{R^*}
\def\Qm{Q_m^*}
\def\GvdW{FSM\xspace}

\usepackage{xcolor}
\newcommand{\red}[1]{{\color{black}#1}}

\shorttitle{Dynamics of thin liquid films on vertical cylindrical fibers}
\shortauthor{H. Ji, C. Falcon, A. Sadeghpour, Z. Zeng, Y. S. Ju, A. L. Bertozzi}

\title{Dynamics of thin liquid films on vertical cylindrical fibers}

\author{H. Ji\aff{1}
  \corresp{\email{hangjie@math.ucla.edu}},
  C. Falcon\aff{1},
    A. Sadeghpour\aff{2},
      Z. Zeng\aff{2},\\
  Y. S. Ju\aff{2}, \and A. L. Bertozzi\aff{1,2}}

\affiliation{\aff{1}Department of Mathematics, University of California, Los Angeles
Los Angeles, CA 90095, USA
\aff{2}Mechanical and Aerospace Engineering Department, University of California, Los Angeles
Los Angeles, CA 90095, USA}

\begin{document}

\maketitle

\begin{abstract}

Recent experiments of thin films flowing down a vertical fiber with varying nozzle diameters present a wealth of new dynamics that illustrate the need for more advanced theory. We present a detailed analysis using a full lubrication model that includes slip boundary conditions, nonlinear curvature terms, and \red{a film stabilization term}. This study brings to focus the presence of a stable liquid layer playing an important role in the full dynamics. We propose a combination of these physical effects to explain the observed velocity and stability of traveling droplets in the experiments and their transition to isolated droplets. This is also supported by stability analysis of the traveling wave solution of the model.

\end{abstract}

\begin{keywords}
\end{keywords}

\section{Introduction}

Thin liquid films flowing down vertical fibers exhibit complex and interesting interfacial dynamics, including the formation of droplets and traveling wave patterns (\cite{quere1990thin,kalliadasis2011falling}).  Such dynamics is an important consideration in various applications (\cite{chinju2000string, zeng2017experimental}) that take advantage of extended interfacial areas afforded by thin liquid films. A recent experimental study (\cite{sadeghpour2017effects}) observed three distinct regimes of interfacial patterns by simply varying the diameters of the nozzles feeding the fluid.  This was quite unexpected because other experimental conditions (flow rate, fiber radius, and fluid), which were thought to primarily govern interfacial dynamics, remained the same. These experimental results motivate us to revisit the existing modeling studies in the literature and extend them for an improved understanding of the physics involved.

The Rayleigh-Plateau instability and the effects of gravity modulation that leads to the rich variety of dynamical behaviors, including the droplet formation, sliding droplets of constant speeds, and irregular waves patterns, have been extensively studied (\cite{kliakhandler2001viscous}). 
A key physical feature of films falling down vertical fibers is that the surface tension plays both a stabilizing and destabilizing role due to the axial and azimuthal curvatures of the interface, respectively (\cite{craster2009dynamics}).
\cite{frenkel1992nonlinear}, \cite{chang1999mechanism} and \cite{kalliadasis1994drop} investigated a weakly nonlinear thin film model under the assumption that the film thickness is much smaller than the fiber radius. These studies reveal that the system can exhibit interesting dynamics with large-magnitude waves. A thick-film (KDB) model was later proposed by \cite{kliakhandler2001viscous}, which utilizes fully nonlinear curvature terms for the case where the film thickness is larger than the fiber radius. However, their model was not derived asymptotically and overestimated the bead velocity. \cite{craster2006viscous} revisited this problem and derived an  asymptotic (CM) model using a low-Bond-number, surface-tension-dominated theory. These two studies focused on cases with relatively small flow rates and developed single evolution equations. To study the case of moderate flow rates, \cite{trifonov1992steady} firstly formulated a system of evolution equations for both the film thickness and volumetric flow rate. This model was then re-formulated by \cite{sisoev2006film} using the integral boundary layer method.   It was more recently revisited by \cite{ruyer2008modelling}, \cite{duprat2009spatial} and \cite{ruyer2012wavy}. 

In \cite{kliakhandler2001viscous}, with decreasing mass flow rates, three different flow regimes were observed from experiments: $(a)$ convective regime where faster moving droplets collide into slower moving ones, $(b)$ Rayleigh-Plateau regime where stable traveling wave propagates without any collisions, and $(c)$ isolated droplet dripping regime where widely spaced traveling beads are separated by smaller droplets. 

Regimes $(b)$ and $(c)$ were qualitatively captured by both the KDB and CM models using traveling wave solutions but the calculated bead velocities were overestimated by more than $40\%$. Moreover, the CM theory led to a conclusion that regime $(b)$ would be a transient rather than a steady-state phenomenon. These discrepancies were further investigated by \cite{duprat2007absolute} and \cite{smolka2008dynamics}.  However, quantitative models that can resolve the reported discrepancies are still lacking, motivating further studies.

In the present paper, we report a combination of experimental and numerical results for flows in the Rayleigh-Plateau regime (regime $(b)$), in particular under flow conditions where the film thickness is comparable to or larger than the fiber radius.  The recent work by \cite{sadeghpour2017effects} showed that one can observe all three flow regimes under a fixed flow rate and a fixed fiber radius by simply varying the nozzle diameter. We leverage their study to examine the characteristics of nonlinear traveling waves that dominate flow dynamics downstream of the nozzle.  We incorporate, into a lubrication model, the slip condition, the fully nonlinear curvature term and a film stabilization  term for the dynamic pressure. We examine their influences on the wave propagation velocity and the dynamic transition from regime $(b)$ to regime $(c)$.

Most of the previous models summarized above employed the classical no-slip boundary condition at the solid-liquid interface.  \cite{haefner2015influence} incorporated slip boundary conditions into a thin-film model and showed that slippage strongly affects the growth rate of undulations when gravitational effects are neglected.  \cite{halpern2017slip} later demonstrated that slip effects promote droplet formation and provided a plausible explanation for the discrepancy between the predicted and experimentally-obtained critical Bond number for droplet formation. More recently, \cite{chao2018dynamics} reported that wall slippage also enhances the size and speed of droplets for thin liquid films flowing down a uniformly heated (cooled) cylinder. All of these past slip models assumed that the liquid film thickness is much smaller than the fiber radius, which is not true in our case.  We quantitatively investigate the slip effects for the first time on flow dynamics where the fluid film thickness is comparable to the fiber radius. 

To describe the wetting behavior of a liquid on a solid substrate, intermolecular forces such as van der Waals interactions and Born repulsion are usually modeled by adding a disjoining pressure in lubrication equations. Different forms of the disjoining pressure representing a combination of long-range and short-range intermolecular forces can be used to characterize hydrophobic or hydrophilic phenomenon. For a well wetting liquid studied in \cite{reisfeld1992non}, these forces favor a thick film, and the disjoining pressure can be described as $\Pi(h) = -A/h^3$ with a positive Hamaker constant $A>0$. \red{In contrast, for a dewetting liquid, the purely destabilizing intermolecular forces modeled by $\Pi(h) = A/h^3$ can lead to finite-time rupture in the film thickness. For partially wetting liquids, a combination of stabilizing and destabilizing molecular interactions are involved and the disjoining pressure takes the form $\Pi(h) = -A/h^3 (1-b/h)$ (\cite{thiele2011depinning}).} 
For an extensive review of this topic, we refer the readers to \cite{RevModPhys.57.827, bonn2009wetting} and \cite{israelachvili2011intermolecular}.  The role of the intermolecular forces in slowly withdrawing a thin fiber out of a bath of wetting liquid has been studied in \cite{quere1989making} and \cite{ quere1999fluid}. The dynamics of non-isothermal liquid film with van der Waals interactions on a horizontal cylinder was considered in \cite{reisfeld1992non} and \cite{thiele2011depinning}. \red{To the best of our knowledge, the stabilization of the coating film in dynamics of liquid flowing down vertical fibers has not been discussed in the literature. In this paper we propose a film stabilization  model to account for thin undisturbed layers of well-wetting silicone oil found in our experiments. }

The stability of the traveling beads plays a key role in the flow regime transition. For the case of thin films of liquid, classical theory developed by \cite{kalliadasis1994drop, chang1999mechanism, yu2013velocity} and experiments by \cite{quere1990thin} show that the behavior of the well-separated pulses is determined by the local undisturbed film thickness connecting these pulses. For the case where the undisturbed film is thicker than a critical value $h_c$, the pulse grows by collecting fluid from the undisturbed liquid layer in between the pulses (regime (c)); while for thinner films, a stable traveling wave propagates at a constant speed (regime (b)). In this study, we focus on the case where the fiber is coated with relatively thick films of liquid compared to the fiber radius, and show that the inclusion of the film stabilization  term allows us to better capture the  undisturbed liquid layer  and thereby the regime transition.

The structure of this paper is as follows. 
Experimental setup and observations in the Rayleigh Plateau regime are presented in section \ref{sec:experiment}.
In section \ref{sec:formulation}, the
model for viscous thin films flowing down a wetting fiber that incorporates wall slippage, nonlinear curvature, and a film stabilization  term is
formulated. 
The traveling wave pattern that appears in the model is examined in section \ref{sec:gvdw}. In this section, we also discuss the influences of the film stabilization term on the moving speed and profile of sliding droplets. In addition, the stability of the spatially uniform solutions and traveling wave solutions for the film stabilization  model is explored. New experimental results, parametrized by varying nozzle diameters, are compared with theory in section \ref{sec:results}. These comparisons reveal that the discrepancy between experiment and theory in the moving speed of droplets, for regimes $(b)$ and $(c)$, can be reasonably resolved by including the film stabilization  term.  Concluding notes and discussion of the remaining open questions are presented in section~\ref{sec:conclusion}.

\section{Experiments}
\label{sec:experiment}
\subsection{Methods}\label{sec:exp_methods}
\begin{figure}
   \centerline{\includegraphics[width = 0.6\textwidth]{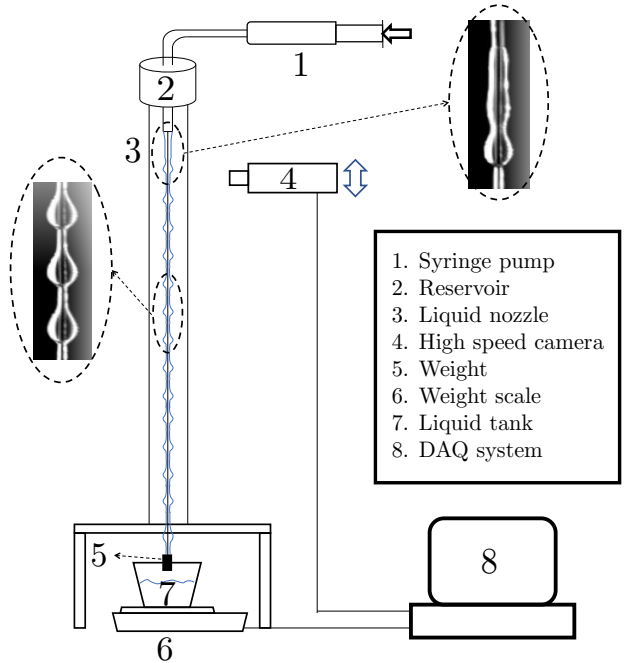}}
   \caption{ Experimental setup.  }
   \label{expsetup}
\end{figure}
Figure \ref{expsetup} shows a schematic of the  setup we used to experimentally study the characteristics of a liquid film flowing down a vertical string. We use a programmable syringe pump to introduce a working liquid into the nozzle and generate flows. The high-speed camera is mounted on a translation stage for focusing and positioning and is operated at a frame rate of 1000 frames/second.

The working fluid is a well-wetting liquid of low surface energy, Rhodorsil silicone oil v50, with density $\rho = 963$ kg/m$^3$, kinematic viscosity $\nu = 50$ mm$^2$/s, surface tension $\sigma = 20.8$ mN/m at \ang{20}C, and capillary length $l_c = 1.5$ mm. We use stainless steel nozzles with the nozzle inner diameters ($ID$s) ranging from $0.6$ mm to $2.5$ mm and the wall thicknesses ranging from $0.1$ mm to $0.2$ mm. The experiments are performed using 0.6 m long Nylon strings of diameter $0.2$ and $0.43$ mm, both smaller than the capillary length. A weight is attached to vertically align the string.  Two X-Y stages are used to center the string with respect to the nozzle. The liquid mass flow rate, monitored using a weight  scale connected to a computer, is varied  from 0.02 g/s to 0.09 g/s.

\subsection{Data Analysis} \label{sec:data_analysis}

\begin{figure} 
      \centerline  { \includegraphics[width = \textwidth]{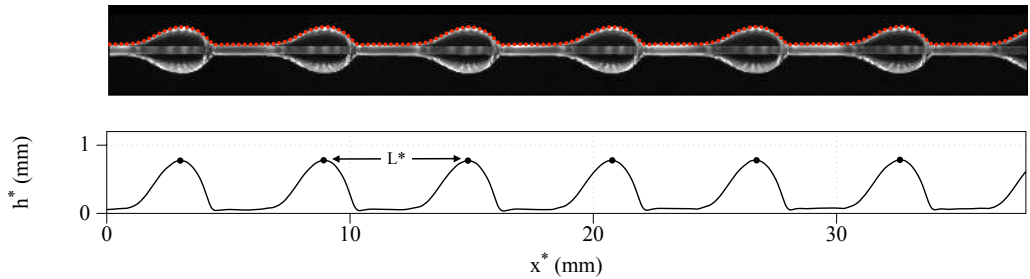} }
   \caption{ Experimental frame (top) and processed data analysis (bottom) for fiber radius $\R = 0.1 \, \si{mm}$, nozzle inner diameter $ID = 0.8$ \si{mm}, and flow rate $\Qm$ = 0.04 \si{g/s}, \red{shown at a distance 8 - 10 mm from the nozzle} . \red{The red dots superimposed on the experimental frame correspond to the extracted film profile and the black dots on the bottom plot} correspond to the locations of the maxima. \red{The average distance between two maxima is $L^* = 5.92$ mm and the film thickness between the drops ranges from 0.095 to 0.120 mm.}} 
   \label{dataanalysis}
\end{figure}

We use image processing tools to extract parameters such as the average bead speed, period $L^*$, maximum film thickness $h_m^*$, and mass contained in each bead. In order to characterize the liquid bead profile, we use the longitudinal distance between the fiber and the maximum curvature point along the liquid. We utilize the color contrast on the image to extract the contour of the fluid and the fiber to thereby determine the liquid film thickness $h^*$. A local least squares smoothing is performed to account for pixelation noise. From this profile, we determine the average period $L^*$ as the distance between two adjacent maxima. Integrating the profile over the chosen domain, we find the mass constraint value $M_0$ as defined later in section \ref{sec:formulation}. Figure \ref{dataanalysis} shows a representative experimental frame and the film profile produced after the data is processed. 

The uncertainties in the parameters obtained from our image processing are estimated in terms of a single pixel scaling and the selection of the color value for the contour. The estimated uncertainty is $\pm0.08$ mm for the streamwise length and $\pm 0.10$ mm for the bead height. Using a total of 1000 images for each run, we measure the averages for maximum fluid height, mass, bead speed, and distance between two subsequent beads.

\section{Model formulation}
\label{sec:formulation}
\begin{figure}
\centering
\includegraphics[width = 0.32\textwidth]{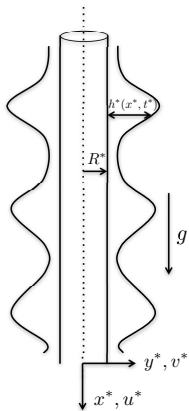}
\caption{Schematic of a thin liquid film flowing down a vertical cylindrical fiber.} 
\label{fig:fiberScheme}
\end{figure}

We consider a flow of two-dimensional axisymmetric Newtonian fluid  down a vertical cylinder of radius $R^*$ (see Figure~\ref{fig:fiberScheme}). The liquid properties, including the surface tension $\sigma$, density $\rho$, and kinematic viscosity $\nu$, are all assumed to be constant. \red{ This model formulation does not include the nozzle size as a system parameter because we are interested in the flow downstream where the nozzle does not affect the dynamics. Instead, we consider the scales for frequency and mass of the droplets, which depend on the nozzle size (\cite{sadeghpour2017effects}). We will discuss this dependence later in Appendix \ref{app:nozzle}. } We review below the derivation of the governing equations and boundary conditions by 
 \cite{ruyer2008modelling} and \cite{craster2006viscous} 
and discuss our inclusion of additional physics related to slip, curvature, and wetting properties.

The dimensional Navier-Stokes equations for axisymmetric flows are
\begin{subequations}\label{fullNS}
\begin{equation}
u^*_{t^*} + u^* u^*_{x^*} + v^*u^*_{y^*} = -\frac{1}{\rho}(p^*_{x^*}+\red{\Pi^*_{x^*}})+g+\nu\left(u^*_{x^*x^*}+\frac{u^*_{y^*}}{y^*}+u^*_{y^*y^*}\right),
\end{equation}
\begin{equation}
v^*_{t^*} + v^* v^*_{y^*} + u^*v^*_{x^*} = -\frac{p^*_{y^*}}{\rho}+\nu\left(\frac{v^*_{y^*}}{y^*} +v^*_{y^*y^*}-\frac{v^*}{y^{*2}}+v^*_{x^*x^*}\right),
\end{equation}
where $t^*$ represents the time, $u^*$ and $v^*$ represents the axial and radial components of the velocity, $p^*$ is the pressure, and $g$ is the gravitational acceleration.  
\red{We adopt the functional form of the disjoining pressure used in \cite{reisfeld1992non} and introduce the film stabilization term $\Pi^*(h^*)$
\begin{equation}
    \Pi^*(h^*) = -\frac{A^*}{{h^*}^3},
\end{equation}
where  $A^*$ is a stabilization parameter.
}
The equation of continuity is given by
\begin{equation}\label{dim_continuity}
v^*_{y^*}+\frac{v^*}{y^*} + u^*_{x^*} = 0.
\end{equation}
Along the fiber, at the interface between the solid substrate and the fluid $y^* = R^*$, we impose the Navier slip and no penetration boundary conditions:
\begin{equation}
v^* = 0, \qquad u^* = \lambda^* u^*_{y^*} \qquad \text{at} \quad y^* = R^*,
\label{slip_bc}
\end{equation}
where $\lambda^* > 0$ is the slip length in standard slip models (\cite{haefner2015influence,munch2005lubrication}). The no-slip boundary condition corresponds to $\lambda^*= 0$. Typical slip lengths for polymeric liquids such as silicone oil range from $1$ to $10  \mu\text{m}$ (\cite{halpern2017slip, quere1990thin}). The normal and shear stress balances on the free surface
$y^* = R^*+h^*$ are given by
\begin{equation}
p^* = \frac{2\mu}{1+h_{x^*}^{*2}}(h_{x^*}^{*2}u^*_{x^*}-h^*_{x^*}(v^*_{x^*}+u^*_{y^*})+v^*_{y^*})+\frac{\sigma}{(1+h_{x^*}^{*2})^{3/2}}\left(\frac{1+h_{x^*}^{*2}}{R^*+h^*}-h^*_{x^*x^*}\right),
\end{equation}
\begin{equation}
(1-h_{x^*}^{*2})(v^*_{x^*}+u^*_{y^*})+2h^*_{x^*}(v^*_{y^*}-u^*_{x^*})=0,
\end{equation}
where $\mu$ is the dynamic viscosity, and $\sigma$ scales the total curvature which consists of a destabilizing azimuthal curvature term and a stabilizing axial curvature term.
We then complete the system by including the kinematic boundary condition on the free surface $y^* = R^*+h^*$,
\begin{equation}
h^*_{t^*} + u^*h^*_{x^*} = v^* \quad \text{at} \quad y^* = R^*+h^*.
\end{equation}
\end{subequations}

The next step is to choose the appropriate dimensionless parameters in order to nondimensionalize the model. Following \cite{duprat2009spatial} we choose the scales for the system as follows: the lengthscale in the radial direction $y$ is $\mathcal{H}$, and the lengthscale in the streamwise direction $x$ is $\mathcal{L} = \mathcal{H}/\epsilon$. The scale ratio $\epsilon$ is set by the balance between the surface tension term $\sigma h^{*}_{x^*x^*x^*}$\red{, arising from $p^*_{x^*}$,} and the gravity $g$, and is given by $\epsilon = (\rho g \mathcal{H}^2/\sigma)^{1/3}$. This scale ratio is small (around $0.4$) in typical experiments and can also be rewritten as $\epsilon = \text{We}^{-1/3}$, where the Weber number $\text{We} = (l_c/\mathcal{H})^2$ compares the capillary length $l_c = \sqrt{\sigma/(\rho g)}$ to the radial lengthscale $\mathcal{H}$. Then the characteristic streamwise velocity is $\mathcal{U}=(g\mathcal{H}^2)/{\nu}$, and
the pressure- and time-scales are given by $\rho g \mathcal{L}$ and $(\nu\mathcal{L})/(g\mathcal{H}^2)$ respectively. 
With these scales, we drop the star superscript in \eqref{fullNS} 
and write the non-dimensional Navier-Stokes equations using dimensionless variables in the form
\begin{subequations}\label{nondim_NS}
\begin{equation}
\epsilon^2\Re\left(u_{t} + u u_{x} + vu_{y}\right) = -p_{x}\red{-\Pi_x}+1+\frac{u_{y}}{y}+u_{yy}+\epsilon^2 u_{xx},
\end{equation}
\begin{equation}
\epsilon^4\Re\left( v_{t} + v v_{y} + uv_{x}\right) = -p_{y}+\epsilon^2\left(\frac{v_{y}}{y} +v_{yy}-\frac{v}{y^{2}}\right)+\epsilon^4 v_{xx},
\end{equation}
where the Reynolds number $\Re = \mathcal{UL}/\nu$. The dimensionless continuity equation is identical to \eqref{dim_continuity} in form. The balances of normal and tangential stresses at $y = h+R$ are expressed as 
\begin{equation}
(1-\epsilon^2 h_x^2)(\epsilon^2 v_x+u_y) + 2\epsilon^2 h_x(v_y-u_x) = 0,
\end{equation}
\begin{align}
p = &\frac{\alpha}{\epsilon^2(1+\alpha h)(1+\epsilon^2 h_x^2)^{1/2}}
- \frac{h_{xx}}{(1+\epsilon^2 h_x^2)^{3/2}}\nonumber\\
&+\frac{\epsilon^4}{1+\epsilon^2 h_x^2}\left[h_x^2 u_x - h_x v_x+ \epsilon^2(-h_x u_y + v_y)\right],
\label{nd_pressure}
\end{align}
where the dimensionless parameter $\alpha = \mathcal{H}/R^*$ is the aspect ratio of the characteristic radial length scale and the fiber radius, and the dimensionless fiber radius $R = R^*/\mathcal{H}$. The slip and no-penetration boundary conditions on $y = R$ are
\begin{equation}
v = 0, \qquad u = \lambda u_y,
\label{nd_slipbc}
\end{equation}
where the non-dimensional slip length $\lambda = \lambda^*/\mathcal{H}$. The kinematic boundary condition remains unaltered in its form.
\end{subequations}

We next simplify the above set of governing equations by following \cite{ruyer2008modelling}. 
Under the lubrication approximation, the inertial contributions can be neglected since $\Re = O(1)$ and $\epsilon\ll 1$. Note that we do not assume the small ratio $\alpha$ between the liquid film thickness and the fiber radius. 
Omitting the terms of order $O(\epsilon^2)$, we rewrite the leading order non-dimensional reduced momentum and continuity equations for the velocity field $(u,v)$ and the dynamic pressure $p$ as
\begin{subequations}
\begin{align}
1 -\frac{\partial p}{\partial x}\red{-\frac{\partial \Pi}{\partial x}}+\frac{\partial^2 u}{\partial y^2} +\frac{u_y}{y}&= 0,\label{px}\\
-\frac{\partial p}{\partial y} &= 0,\label{py}\\
\frac{\partial u}{\partial x}+\frac{\partial v}{\partial	y} +\frac{v}{y} &= 0.\label{incomp}
\end{align}
The balance of tangential stresses at the free surface $y = h + R$ is reduced to
\begin{equation}
u_y = 0.
\label{shear}
\end{equation}
The destabilizing azimuthal curvature and the stabilizing streamwise curvature terms in \eqref{nd_pressure} are both important throughout the liquid film. For $\epsilon \ll 1$ a formal expansion of the azimuthal curvature term in  \eqref{nd_pressure} yields
$$
 \frac{\alpha}{\epsilon^2(1+\alpha h)\sqrt{1+\epsilon^2 h_x^2}} = \frac{\alpha}{\epsilon^2(1+\alpha h)}
 \left(1-\frac{1}{2}\epsilon^2 h_x^2+O(\epsilon^4)\right),
$$
which shows that this term is of order  $O(\alpha/\epsilon^2)$, for both cases $\alpha h \ll 1$ and $\alpha h = O(1)$. 
At the steep front of a moving droplet, we have $h_x = O(1)$, and the second term in the expansion contributes an additional $O(\alpha)$ term to the dynamic pressure.
Similarly, for the streamwise curvature term in \eqref{nd_pressure}, we have
$$
\frac{h_{xx}}{(1+\epsilon^2 h_x^2)^{3/2}} = {h_{xx}}\left(1-\frac{3}{2}\epsilon^2 h_x^2 + O(\epsilon^4)\right).
$$
At the front of the moving droplet, the $O(\epsilon^{2})$ term introduced by the nonlinear curvature is negligible. Therefore, we only keep a fully nonlinear azimuthal curvature term and use the linearized curvature in the streamwise direction.
The balance of normal stresses at the free surface $y = h + R$ is then reduced to
\begin{equation}
p = \frac{\alpha}{\epsilon^2(1+\alpha h)\sqrt{1+\epsilon^2 h_x^2}}-\frac{\partial^2 h}{\partial x^2}.
\label{pressure}
\end{equation}
\end{subequations}
It is important to note that the first term on the right-hand-side of \eqref{pressure} accounts for the balance between the azimuthal and axial scales characterized by $\alpha$ and $\epsilon$. We will further discuss appropriate forms of this term for different cases later.

To derive the evolution equation for $h$ from above, we first consider a uniform Nusselt flow without any interfacial instabilities. The velocity field of this flow is obtained by balancing the viscosity and gravity acceleration in \eqref{nondim_NS}. 
That is, the streamwise velocity $u_0$ of the Nusselt flow satisfies the reduced system
\begin{subequations}\label{nusselt}
\begin{equation}
1+\frac{\partial^2 u_0}{\partial y^2} + \frac{1}{y}\frac{\partial u_0}{\partial y} = 0
\end{equation}
with boundary conditions
\begin{equation}
\frac{\partial u_0}{\partial y} = 0 \text{ at } y=h+R, \qquad u_0 = \lambda \frac{\partial u_0}{\partial y} \text{ at } y=R.
\end{equation}
\end{subequations}
Solving \eqref{nusselt} for $u_0$ leads to 
\begin{equation}
u_0(y) = -\frac{1}{4}(y^2-R^{2})+\frac{1}{2}(h+R)^2\ln\left(\frac{y}{R}\right)+h\lambda\left(\frac{h}{2R}+1\right).
\label{eq:u0}
\end{equation}
The rescaled kinematic boundary conditions
together with the slip and no-penetration boundary conditions \eqref{nd_slipbc} and the shear stress condition \eqref{shear}
lead to the mass conservation equation
\begin{equation}
(1+\alpha h)\frac{\partial h}{\partial t} + \frac{\partial q}{\partial x} = 0, \text{  where } q = \frac{1}{R}\int_{R}^{h+R} uy~dy.
\label{massconserv}
\end{equation}
Following \cite{ruyer2008modelling} \red{where an approach based on a projection of the velocity field of a test function is used,}
we take the inner product of \eqref{px} with the Nusselt uniform solution $u_0$, and obtain
\begin{equation}
\int_{R}^{h+R}(1+u_{yy}+\frac{u_y}{y})u_0 y~dy = \int_{R}^{h+R} (p_x\red{+\Pi_x}) u_0 y~dy.
\end{equation}
From equation \eqref{py}, we see that the pressure $p$ is a function of $x$ only, and the right-hand-side of the above equation becomes $p_x \bar{R}q_0$. Note that the Nusselt solution $u_0$ satisfies \eqref{nusselt}. By integrating by parts and applying the slip and no-penetration boundary conditions \eqref{nd_slipbc} and the shear stress condition \eqref{shear}, we obtain 
\begin{equation}\label{q}
q = \left(1-\frac{\partial p}{\partial x}\red{-\frac{\partial\Pi}{\partial x}}\right)q_0.
\end{equation}
Here $q_0$, the flow rate per unit circumference length for the uniform Nusselt layer, is
\begin{equation}\label{nusselq}
q_0 = \frac{1}{R}\int_{R}^{R+h} u_0 y ~dy =\frac{h^{3}}{3}\phi\left(\alpha h\right)+\frac{h^{2}}{4}\left(\alpha h+2\right)^2\lambda,
\end{equation}
\red{where the shape factor $\phi$ is a function defined by
\begin{equation}
\phi(X) = \frac{3}{16X^3}[(1+X)^4(4\log(1+X)-3)+4(1+X)^2-1].
\end{equation}
}
Given a dimensional volumetric flow rate $Q_m^*$ and fiber radius $R^*$, we define the volumetric flow rate per circumference unit $q_0^*$ as $q_0^* = Q_m^*/(2\pi \rho R^*)$. From \eqref{nusselq} the characteristic axial lengthscale $\mathcal{H}$ for a uniform Nusselt flow can then be obtained. For the no-slip case $\lambda=0$, equation \eqref{nusselq} gives the Nusselt solution $h \equiv \mathcal{H}$ used in \cite{duprat2009spatial,ruyer2008modelling}.

Finally, by rescaling  the time scale $t \to {t}/{\phi(\alpha)}$ and combining \eqref{massconserv},
\eqref{nusselq} and \eqref{q}, we obtain the dimensionless governing equation
\begin{equation}\label{eq:main_1}
\frac{\partial}{\partial t} \left(h+\frac{\alpha}{2}h^2\right) + \frac{\partial}{\partial x}\left[\mathcal{M}(h)\left(1-\frac{\partial p}{\partial x}\red{-\frac{\partial \Pi}{\partial x}}\right)\right] = 0,
\end{equation}
where the mobility function takes the form
\begin{equation}\label{mobility}
\mathcal{M}(h;\lambda, \alpha) = \frac{h^3}{3}\frac{\phi(\alpha h)}{\phi(\alpha)}+\frac{h^2(\alpha h+2)^2\lambda}{4\phi(\alpha)}.
\end{equation}
This choice of timescale leads to a normalized mobility function such that $\mathcal{M} = 1/3$ for $h = 1$ and $\lambda = 0$.
Moreover, in the limit of the aspect ratio $\alpha\to 0$, the mobility function becomes
\begin{equation}
\mathcal{M}(h;\lambda, \alpha)  = \frac{1}{3}h^2(h+3\lambda)+\frac{\alpha}{3}h^2(h+3\lambda)(h-1)+O(\alpha^2),
\end{equation}
and its leading order term agrees with the mobility function used in the slip model proposed by \cite{halpern2017slip}. Substituting \eqref{pressure} into \eqref{eq:main_1} yields an evolution equation for the film thickness $h(x,t)$ 
\begin{equation}
\label{eq:main}
\boxed{\hspace{12pt}
\frac{\partial}{\partial t} \left(h+\frac{\alpha}{2}h^2\right) + \frac{\partial}{\partial x}\left[\mathcal{M}(h)\left(1-\frac{\partial }{\partial x}\left[\mathcal{Z}(h)-\frac{\partial^2 h}{\partial x^2}\right]\right)\right] = 0,\hspace{12pt}}
\end{equation}
where $\mathcal{Z}(h)$ includes the destabilizing azimuthal curvature of the film and the \red{film stabilization term}. 
\red{Comparing \eqref{eq:main} and \eqref{eq:main_1},  and using \eqref{pressure} with a scaling parameter $\eta = \epsilon^2$,  we write the complete form of $\mathcal{Z}(h)$ as
$$
\mathcal{Z}(h) = p+h_{xx} + \Pi = \frac{\alpha}{\eta(1+\alpha h)\sqrt{1+\eta h_x^2}} + \Pi(h).
$$}
Equation \eqref{eq:main} is a fourth-order nonlinear partial differential equation for the thickness $h(x,t)$. This model accounts for the surface tension, gravity and azimuthal instabilities, but neglects inertia and streamwise viscous dissipation (\cite{ruyer2009film}). When the Rayleigh-Plateau instability dominates over the inertia and streamwise viscous dissipation, our model \eqref{eq:main} is expected to provide a good agreement with experimental data.

Different forms of $\mathcal{Z}(h)$ appear in the literature to represent the azimuthal curvature of the film. For instance, \cite{yu2013velocity} assumed that the film is much thinner than the fiber radius, $\mathcal{H} \ll R^*$ (which corresponds to $\alpha \ll 1$), and used a simple form $\mathcal{Z}(h) = -(\alpha/\epsilon)^2 h$ where the leading-order constant term in the expansion of the azimuthal curvature is neglected. For thicker films that are of interest to the present study, $\alpha = O(1)$ and $\alpha h$ is at most $O(1)$ so we use \eqref{z1} as the first form of $\mathcal{Z}$
\begin{equation}
\mathcal{Z}_{CM}(h) = \frac{\alpha} {\eta (1+\alpha h)}
\label{z1}
\end{equation}
following \cite{craster2006viscous, sisoev2006film, ruyer2008modelling, ruyer2009film}. 
With $\mathcal{Z_{CM}}(h)$ in \eqref{z1} and $\lambda=0$, the evolution equation \eqref{eq:main} reduces to
\begin{equation}\label{eq:mcm} \textbf{(CM)} \hspace{12pt}
\frac{\partial}{\partial t} \left(h+\frac{\alpha}{2}h^2\right) + \frac{\partial}{\partial x}\left[\frac{h^3}{3}\frac{\phi(\alpha h)}{\phi(\alpha)}\left(1-\frac{\partial }{\partial x}\left[\frac{\alpha}{\eta (1+\alpha h)}-h_{xx}\right]\right)\right] = 0,
\end{equation}
which
is consistent with the evolution equation derived by \cite{craster2006viscous} except for a scaling difference. For the rest of this paper, we will refer to the model \eqref{eq:mcm} as the Craster \& Matar (CM) model. 

To incorporate the slip effects under the framework of the CM model, we use the evolution equation \eqref{eq:main} with the non-dimensional slip length $\lambda > 0$, the mobility function  \eqref{mobility} and the azimuthal curvature term  \eqref{z1}.  This will be referred to as the Slip Craster \& Matar (SCM) model. We will show that this slip model promotes droplet formation and leads to an increased speed of propagation.

With the aspect ratio $\alpha $ being of $O(1)$, in the limit as $\eta \to 0$, \eqref{z1} becomes singular. To address this problem, in this paper we will consider two additional forms of $\mathcal{Z}(h)$ to reflect the balance between the azimuthal and axial scales under different flow conditions. 

As the second form of $\mathcal{Z}$, we define
\begin{equation}
\mathcal{Z}_{FC}(h) = \frac{\alpha} {\eta (1+\alpha h)\sqrt{1+\eta h_x^2}}.
\label{z2}
\end{equation}
This fully nonlinear azimuthal curvature term \red{provides} 
an \red{$O(\alpha/\eta^{3/2})$} contribution 
to the dynamic pressure near the advancing edge of the moving beads. \red{It has been shown in many applications (\cite{snoeijer2006free, lopes2018multiple}) that using the full expression for the curvature term can provide better accuracy for lubrication models, and this issue was recently reviewed in \cite{thiele2018recent}.}
We will show that the inclusion of the fully nonlinear curvature yields an increased speed of propagation of the traveling beads and partially
improves agreement with our experimental results. Compared with the thick film model proposed in \cite{kliakhandler2001viscous}, where both curvature terms are fully nonlinear, our model has only a nonlinear azimuthal curvature term.  The stabilizing axial curvature term is linearly approximated.
For the remainder of this paper, we will refer to the evolution equation \eqref{eq:main} with the mobility function \eqref{mobility} and the azimuthal curvature term \eqref{z2} as the Full Curvature Model (FCM).

As the third form of $\mathcal{Z}$, we include the film stabilization  term to deal with the unbalanced azimuthal curvature term. 
Thus we replace \eqref{z1}
with
\begin{equation}
\mathcal{Z}_{FS}(h) = \frac{\alpha} {\eta (1+\alpha h)} - \frac{A}{h^3}.
\label{z3}
\end{equation}
\red{The last term of \eqref{z3} is motivated by the functional form of the long range attractive part of the well-known apolar van der Waals model} (\cite{RevModPhys.57.827, RevModPhys.69.931, oron2001dynamics}) for the well-wetting liquids used in our experiments.  The stabilization parameter $A = (\alpha^2 \epsilon_p^4)/(3\eta (\alpha \epsilon_p+1)^2) >0$ is expressed in terms of a dimensionless thickness $\epsilon_p$, below which a thin uniform fluid layer \red{on a fiber} is stable.  We will discuss this stability criterion further in section \ref{sec:gvdw}.
The \red{film stabilization term} is found to significantly improve agreement with our experimental data as discussed in section \ref{sec:results}. The model \eqref{eq:main} with $\mathcal{M}(h)$ from \eqref{mobility} and the azimuthal curvature with film stabilization term \eqref{z3} will be referred to as the film stabilization  model (\GvdW).

\begin{table}
\begin{center}
\begin{tabular}{p{6.8cm} p{1.5cm} p{5cm}}
 (CM) Craster \& Matar Model: & $\lambda = 0$ &$\mathcal{Z}(h) = \mathcal{Z}_{CM}(h)$ in \eqref{z1}\\
 (SCM) Slip Craster \& Matar Model: & $\lambda > 0$ &$\mathcal{Z}(h) = \mathcal{Z}_{CM}(h)$ in \eqref{z1}\\
 (FCM) Full Curvature Model: & $\lambda \ge 0$ &$\mathcal{Z}(h) = \mathcal{Z}_{FC}(h)$ in \eqref{z2}\\
 (\GvdW) Film stabilization  model: & $\lambda \ge 0$ & $\mathcal{Z}(h) = \mathcal{Z}_{FS}(h)$ in \eqref{z3}
\end{tabular}
\end{center}
\caption{A summary of the models using \eqref{eq:main}}
\label{table:models}
\end{table}

In summary, this paper considers four versions of the model \eqref{eq:main} listed in Table \ref{table:models} to incorporate different physical effects.
When contrasted with the CM model, which corresponds to the asymptotic model studied in \cite{craster2006viscous}, the other three new models take into considerations the slip effects, the nonlinear curvature, and the film stabilization term, respectively.

\section{ Film stabilization model and stability analysis}
\label{sec:gvdw}

In this section, we derive the stabilization parameter $A$ in equation \eqref{z3} and study its effects. Using a stability analysis study, we show that $A>0$ is important to reproduce the stable traveling wave solutions (TWS) observed in experiments. 

We begin by examining the linear stability of the uniform Nusselt solution.
Following the approach in \cite{craster2006viscous}, we perturb the uniform film ($ h_0 =1$) by an infinitesimal Fourier mode,
\begin{equation}
h = h_0 + \delta e^{\mi k x + \Lambda t },
\label{const_perturbation}
\end{equation}
where $k$ is the wave number, $\Lambda$ describes the growth rate of the perturbation, and $\delta \ (\ll 1)$ the initial amplitude. Expanding PDE \eqref{eq:main} then gives the dispersion relation:
\begin{equation}\label{eq:stability1}
\Lambda = -\mi k c_k(\alpha, \lambda) 
+ \frac{k^2}{3(1+\alpha)}\left(\frac{\alpha^2}{(1+\alpha)^2\eta} - k^2 - 3A\right)\left(1+\frac{3(\alpha+2)^2 \lambda}{4\phi(\alpha)}\right),
\end{equation}
where $c_k$ is the speed of linear kinematic wave solutions of \eqref{eq:main} for small wave numbers,
\begin{equation}
 c_k(\alpha, \lambda)  = \frac{1}{\alpha+1}\left(1+\frac{\alpha\phi'(\alpha)}{3\phi(\alpha)}\right)+\frac{(\alpha+2)\lambda}{\phi(\alpha)},
\end{equation}
and it increases linearly with the presence of slip.
 The real part of $\Lambda$ is the effective growth rate, and 
the thickness $h_0=1$ is long-wave unstable with respect to perturbations with $0 < k < k_c$, where the critical wavenumber 
$
k_c = \left(\frac{\alpha^2}{(1+\alpha)^2\eta}-{3A}\right)^{1/2}.
$
For $A = 0$ this cut-off wavenumber
corresponds to the classical RP mode for the capillary instability of a viscous jet where $\Re(\Lambda)=0$ at $k = k_c$.
The most unstable mode occurs at 
$
k_m = \left(\frac{\alpha^2}{2(1+\alpha)^2\eta}-\frac{3A}{2}\right)^{1/2},
$
 where the largest growth rate $\Lambda_m$ is attained.
The equation \eqref{eq:stability1} also shows that the slip effects always enhance the Rayleigh-Plateau instability which agrees with the work by \cite{halpern2017slip}.

\begin{figure} 
   \centering
   \includegraphics[width=2.6in]{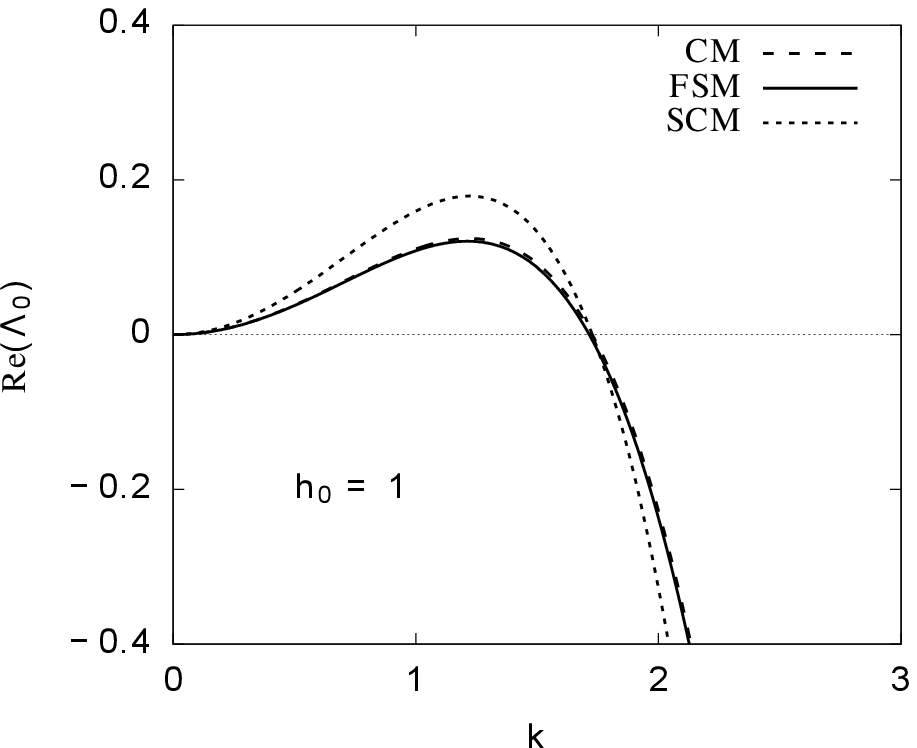} 
      \includegraphics[width=2.6in]{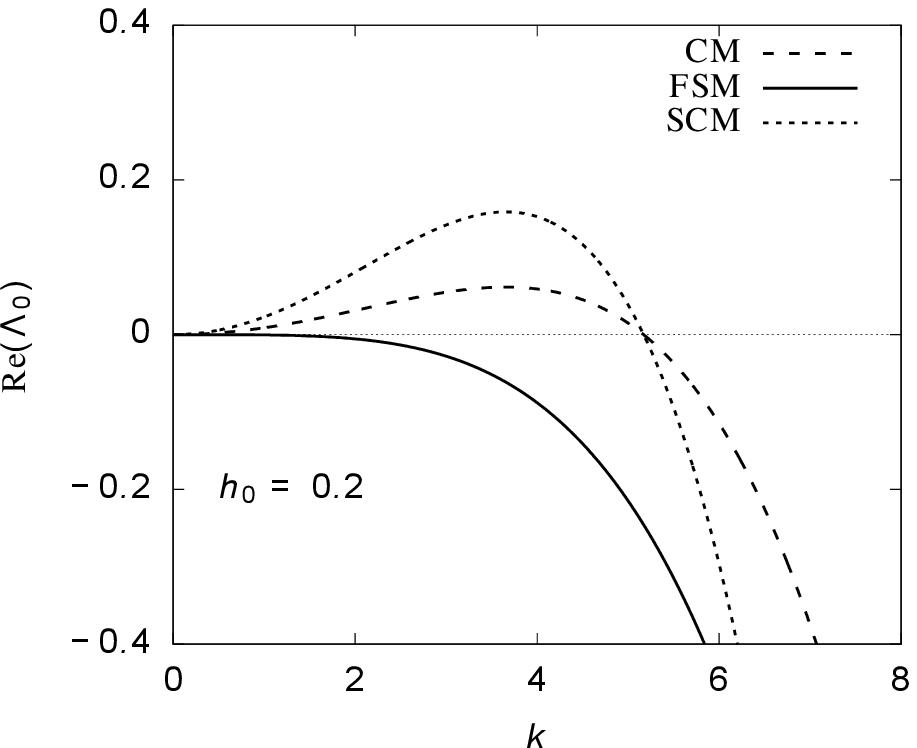} 
   \caption{Dispersion relation plots of $\Re(\Lambda_0)$ against the wavenumber $k$ with the constant film thickness (left) $h_0 = 1$ and (right) $h_0 = 0.2$ for the CM model, the SCM with slip $\lambda = 0.1$, and the \GvdW model with  $\epsilon_p = 0.2$, which corresponds to $A = 0.027$ from equation \eqref{eq:A}. Other system parameters are $\alpha = 4.96$ and $\eta = 0.23$.}
   \label{fig:dispersion}
\end{figure}

We repeat the above analysis for a uniform thin layer $h = h_0  (<1)$. 
The effective growth rate of disturbances with wavenumber $k$ is given by
\begin{equation}
\Re(\Lambda_0) = \frac{k^2}{3h_0(\alpha h_0+1)}\left(\frac{h_0^4 \alpha^2}{(\alpha h_0 +1)^2 \eta}-k^2 h_0^4  -3A\right)\left(\frac{\phi(\alpha h_0)}{\phi(\alpha)}+\frac{3(\alpha h_0 +2)^2}{4h_0\phi(\alpha)} \lambda\right).
\label{dispersion}
\end{equation}
Figure~\ref{fig:dispersion} gives the dispersion relation \eqref{dispersion} for the uniform Nusselt solution $h_0 = 1$ and a thinner undisturbed layer of $h_0 = 0.2$ under the CM, SCM, and \GvdW models. The Nusselt solution $h_0 = 1$ is unstable for small wavenumbers for all three models. The finite slippage increases the growth rates of the unstable modes.  For the CM and SCM models, the thinner undisturbed liquid layer ($h_0 = 0.2$) is  unstable over larger ranges of the wavenumber.  In stark contrast, the \GvdW model  saturates the unstable modes, rendering the thinner undisturbed layer linearly stable for all wavenumbers (see the solid curve in Figure~\ref{fig:dispersion} (right)) for certain values of $A$.

In the literature, $A$ is typically expressed \red{as a Hamaker constant} in terms of microscopic quantities (\cite{RevModPhys.57.827}) . We instead choose the values of $A$ based on stable undisturbed liquid layers. \red{Empirical observations of the thin film layers between droplets  indicate that these} undisturbed liquid layers are comparable to the corresponding fiber radius $R^*$. Therefore, we pick a coating thickness $\epsilon_p^* \approx R^*$ \red{, obtained from experimental measurements}.
Using the dimensionless undisturbed layer thickness $\epsilon_p = \epsilon_p^*/\mathcal{H}$ and the dispersion relation in \eqref{dispersion}, we derive a formula for $\tilde{A}$,
\begin{equation}
\tilde{A} = \frac{\alpha^2 \epsilon_p^4}{3\eta(\alpha\epsilon_p + 1)^2}.
\label{eq:A}
\end{equation}
For $A = \tilde{A}$, any thin flat film of thickness less than the threshold value $\epsilon_p$ is linearly stable, i.e. $\Re(\Lambda_0)< 0$, for all wave numbers. 

We next examine the influences of  the film stabilization  terms  \eqref{z3} on the  profiles and speeds of propagation of the traveling wave patterns governed by \eqref{eq:main}.  
We consider the model over a periodic domain $0 \le x \le L$ \red{and introduce a change of variables to the reference frame of the traveling wave,}
\begin{equation}
\xi = x - ct, \qquad s = t, \qquad h(x,t) = \tilde{h}(\xi, s),
\nonumber
\end{equation}
where $c$ is the speed of the traveling wave. Then $\tilde{h}(\xi,s)$ satisfies the PDE
\begin{equation}
\frac{\partial }{\partial s}\left(\tilde{h}+\frac{\alpha}{2}\tilde{h}^2\right)-c\frac{\partial }{\partial \xi}\left(\tilde{h}+\frac{\alpha}{2}\tilde{h}^2\right)+\frac{\partial}{\partial \xi}\left[\mathcal{M}(\tilde{h})\left(1-\frac{\partial}{\partial \xi}\left[\mathcal{Z}(\tilde{h})-\tilde{h}_{\xi\xi}\right]\right)\right]=0.
\label{eq:travelingPDE}
\end{equation}
\red{The traveling wave solution $H(\xi)$ is a steady state of the PDE \eqref{eq:travelingPDE} and satisfies the fourth-order ordinary differential equation}
\begin{equation}
c\frac{d}{d \xi}\left(H+\frac{\alpha}{2}H^2\right)=\frac{d}{d \xi}\left[\mathcal{M}(H)\left(1-\frac{d}{d \xi}\left[\mathcal{Z}(H)-H_{\xi\xi}\right]\right)\right].
\label{eq:travelingODE}
\end{equation}
This is a nonlinear eigenvalue problem, where the speed of propagation $c$ corresponds to the eigenvalue.
The effects of slippage $\lambda$ and full nonlinear curvature \eqref{z2} will be studied in section \ref{sec:slipCurvature}.
Newton's method is used to solve the nonlinear ODE \eqref{eq:travelingODE} where the speed $c$ is treated as an unknown variable. 
In order to achieve local uniqueness, we impose a constraint of mass conservation
\begin{equation}
\int_0^L H+\frac{\alpha}{2}H^2~d\xi = M_0
\label{eq:mass}
\end{equation}
for given values of mass $M_0$ and wavelength $L$ extracted from our experimentally obtained liquid film profiles.
We also set $H(\xi_0) = H_0$ for $0\le \xi_0 \le L$. 

\begin{figure} 
\centering
\includegraphics[width=2.6in]{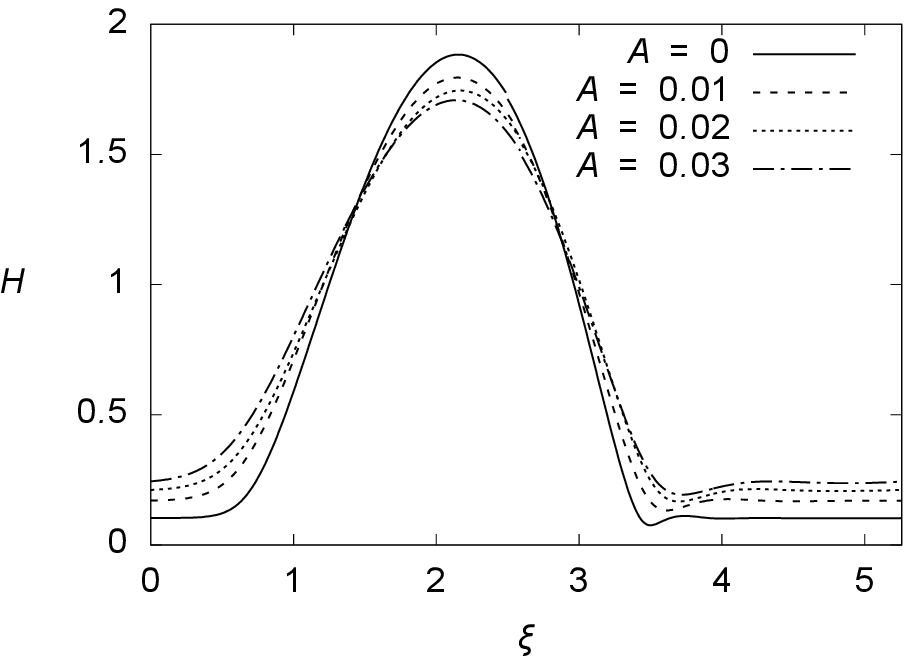}
\includegraphics[width=2.6in]{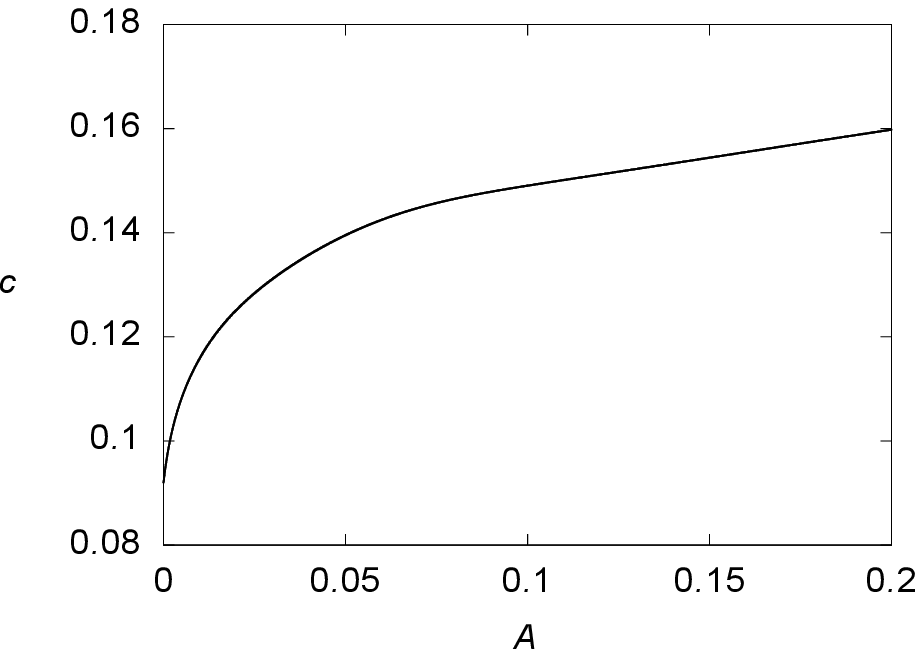}
   \caption{Traveling wave solutions of \eqref{eq:travelingODE} with a varying stabilization parameter $A = 0, 0.01, 0.02, 0.03$ showing that (left) stronger regularization yields smaller droplet heights and thicker precursor layers (right) the moving speed of drops is significantly increased with larger values of $A$.}
   \label{vanDerWaals_comparison}
\end{figure}

Numerical investigations reveal that including the film stabilization  term in \eqref{z3} enhances the moving speed of traveling wave solutions. In Figure~\ref{vanDerWaals_comparison} (left), we plot the profiles of traveling wave solutions with a varying stabilization parameter $A$ and identical mass constraint $M_0$ and period $L$. Since a larger value of $A$ corresponds to a stronger wetting potential in the lubrication model, the near-flat coating thickness increases with increasing $A$, as expected. Correspondingly, the height of the beads decreases as the coating thickness increases under the fixed overall mass constraint. That is, the presence of the film stabilization  term saturates capillary instabilities and generates smaller moving beads.
Figure~\ref{vanDerWaals_comparison} (right) shows that the predicted velocity $c$ increases with the parameter $A$ superlinearly for small $A$, and increases linearly as $A$ becomes larger.
In section \ref{sec:correction_model}, we will show that, compared to the CM model \eqref{eq:mcm}, the \GvdW model with parameter $A$ given by \eqref{eq:A} significantly improves predictions of the traveling wave velocity against experimental observations.

Earlier works have not reported a detailed stability analysis of the traveling wave. Therefore, we take a closer look at the linear stability of the traveling wave solutions in the film stabilization  model.  
We consider a positive periodic traveling wave solution $H(\xi)$ over the domain $0 \le \xi \le L$,
and perturb it by setting $\tilde{h}(\xi,s) = H(\xi)+\delta\Psi(\xi)e^{\Lambda s}$, 
where $\delta \ll 1$ and 
$\Psi(\xi)$ is also L-periodic. \red{Here we only focus on perturbations of the same period since the dynamics in both Rayleigh Plateau and isolated droplet regimes have fixed spatial periods. More complicated droplet dynamics (\cite{kalliadasis1994drop}) in the convective regime is not investigated here.}
We linearize the equation \eqref{eq:travelingPDE} around the steady state $H(\xi)$ and obtain the $O(\delta)$ equation
\begin{equation}
\Lambda \Psi = \mathscr{L} \Psi,
\label{eq:ev_tws}
\end{equation}
where the linear operator $\mathscr{L}$ is
\begin{multline}
\mathscr{L}\Psi \equiv c\left(\frac{d}{d\xi}+\frac{\alpha H'}{1+\alpha H}\right)\Psi + \frac{1}{1+\alpha H}\frac{d}{d\xi}\left(\mathcal{M}(H)\frac{d}{d\xi}\left[\mathcal{Z}'(H)\Psi-\Psi''\right]\right) \\
- \frac{1}{1+\alpha H}\frac{d}{d\xi}\left(\mathcal{M}'(H)\left[1-\frac{d}{d\xi}(\mathcal{Z}(H) - H'')\right]\Psi\right).
\end{multline}
Here $\Psi(\xi)$ is a normalized eigenmode associated with an eigenvalue $\Lambda$ and $\|\Psi\|_2 = 1$. 
If there are any eigenvalues $\Lambda$ with $\Re(\Lambda)>0$ to the problem \eqref{eq:ev_tws}, then the periodic traveling wave solution $H(\xi)$ is unstable. For simplicity we only focus on the case where a one-period solution fits in the domain, and numerically calculate the spectrum and corresponding eigenfunctions for the eigenproblem. 
\begin{figure} 
\centering
\includegraphics[width=2.6in]{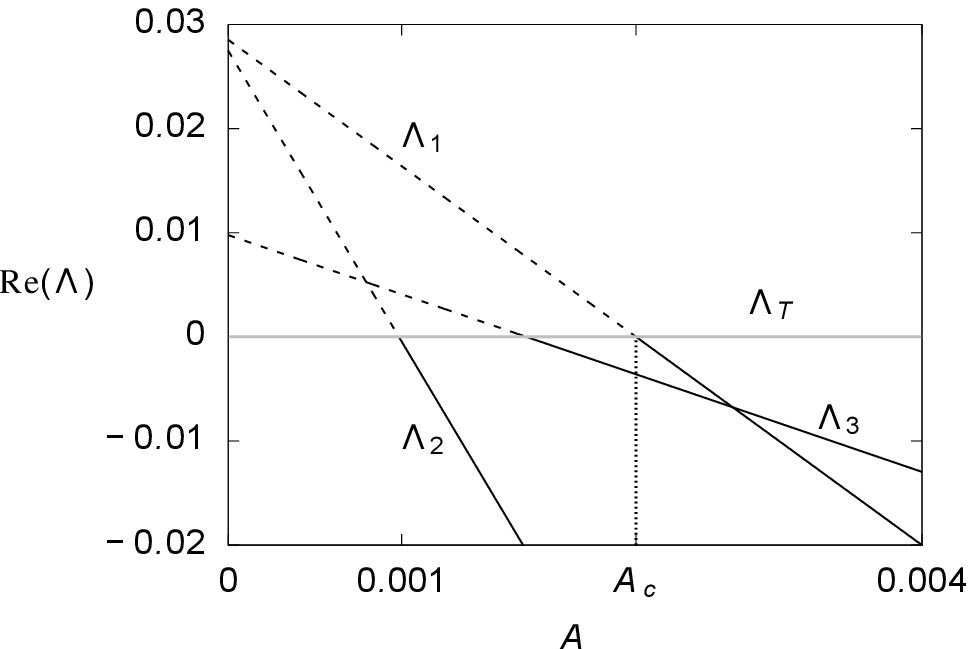}
\includegraphics[width=2.6in]{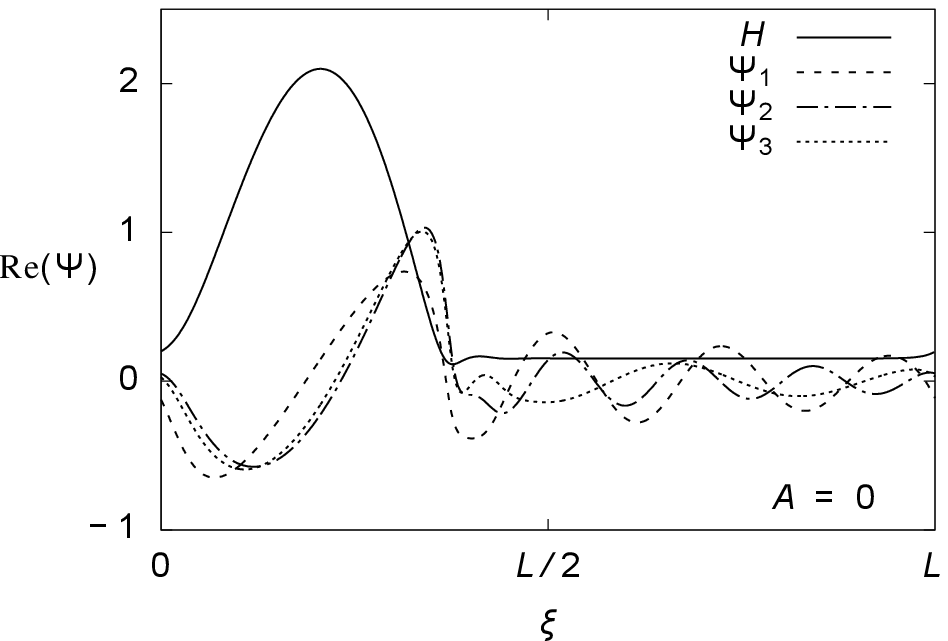}
\caption{(Left) The dependence of dominant eigenvalues of $H(\xi)$ on the stabilization parameter $A$ for the eigenproblem \eqref{eq:ev_tws} where the unstable eigenvalues are plotted in dashed lines and the stable eigenvalues are in solid lines (Right) Corresponding traveling wave profile $H$ and unstable eigenmodes $\Psi_1$, $\Psi_2$, $\Psi_3$ for the CM model ($A=0$). The parameters $(L,M_0) = (8.86,22.16)$ correspond to the experiment with flow rate $Q_m^*=0.04$ g/s, fiber radius $R^*=0.1$ mm, and nozzle diameter $1.06$ mm.}
\label{stability_tws}
\end{figure}

 Figure~\ref{stability_tws} (right) shows the unstable modes predicted by the CM model ($A$ = 0). These instabilities contradict the stable TWS from the corresponding experiment and generate small wavy patterns in the flat film connecting the moving beads. However, these instabilities can be saturated by the \GvdW model with an appropriate stabilization parameter $A$. For a typical traveling wave solution $H(\xi)$ to the \GvdW model, its dominant eigenvalues with a varying $A$ are plotted in Figure~\ref{stability_tws} (left). It shows that without the film stabilization  term (i.e. $A=0$), the TWS is unstable. In addition to the translational eigenmode $\Lambda_T=0$, the dominant unstable eigenvalues are given by 
 \red{complex-conjuagte pairs $\Lambda_1 = 0.0283 \pm 0.756\mi$, $\Lambda_2 = 0.0264 \pm 1.006 \mi$, and $\Lambda_3=0.0097\pm 0.503\mi$. }
\red{Increasing the parameter $A$ yields three pairs of complex conjugate eigenvalues crossing the imaginary axis, suggesting three Hopf bifurcations occur at these crossing points. At each of these bifurcation points, a branch of time-periodic solution emerges corresponding to typical dynamics in the isolated droplet regime.}
 
For $A>A_c$ all the eigenvalues of $H(\xi)$ satisfy $\Re(\Lambda)<0$ and
the traveling wave is stabilized. 
In section \ref{sec:regime_transition}, we will show that the film stabilization term helps better arrest the stability transition from the Rayleigh-Plateau regime to isolated droplet regime in experiments with different nozzle diameters.

\section{Results}
\label{sec:results}

\subsection{Experimental comparisons}
\label{sec:correction_model}

Here, we present a comparison between physical experiments and two models, the film stabilization  (\GvdW) model and the Craster \& Matar (CM) model. Appendix \ref{app:Nomenclature}, shows the range of nozzle sizes, fiber radius, and flow rates for the experiments. For each experimental condition, we extract a characteristic period $L^*$ and bead liquid mass $M_0^*$. We then use these as input parameters in our models.  Appendix \ref{app:nozzle} describes how a unique traveling wave solution (TWS) is selected to compare with each experiment for both the \GvdW and CM models.

\begin{figure} 
   \centerline{
   \includegraphics[width=2.5in]{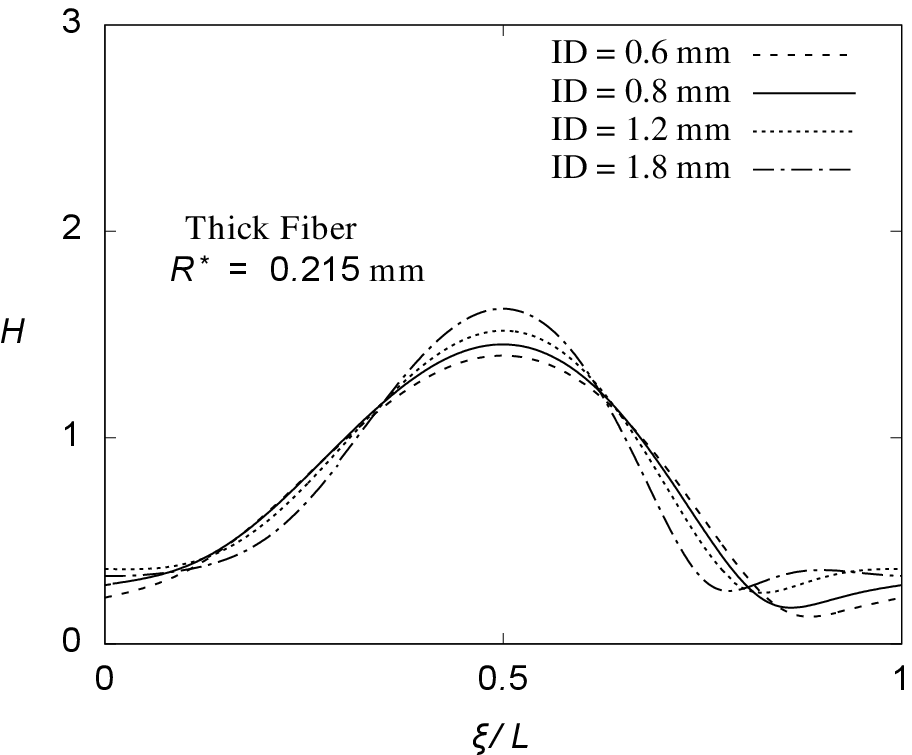} 
   \includegraphics[width=2.5in]{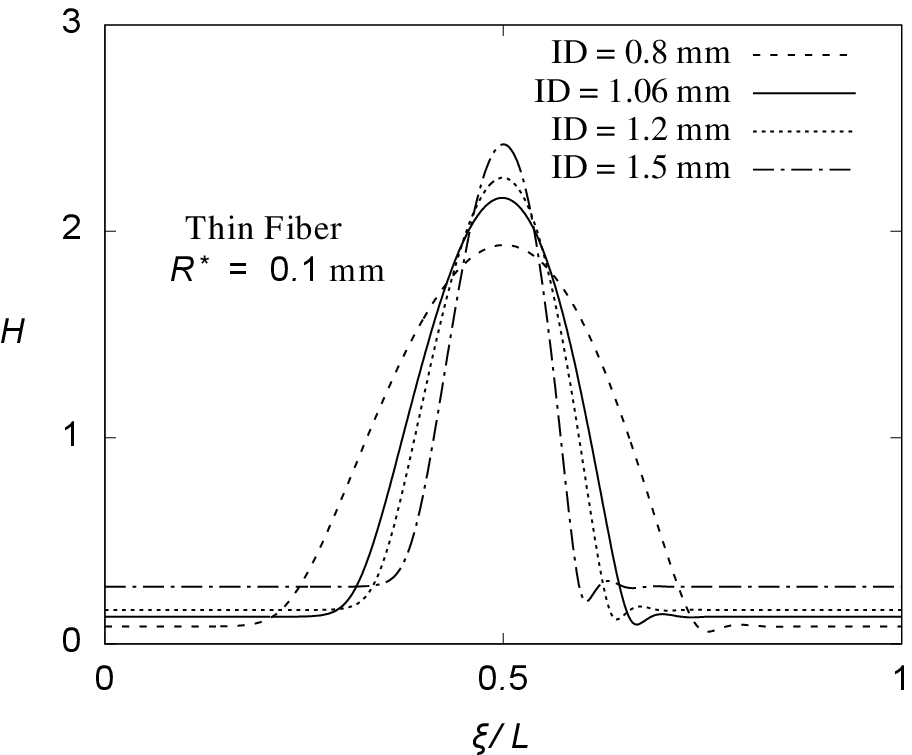}
   }
   \caption{Plots of rescaled traveling wave solutions $H(\xi)$ obtained from \eqref{eq:main} using $(L, M_0)$ from experiments at flow rate $\Qm = 0.04$ g/s and (left) thick fiber of radius $\R = 0.215$ mm (right) thin fiber of radius $\R = 0.1$ mm, corresponding to different nozzle inner diameters (ID). The slip length is set to be $\lambda^* = 7$ $\mu$m.}
   \label{fig:profile_thin_thick}
\end{figure}

\begin{figure} 
   \centering
   \includegraphics[width=1\textwidth]{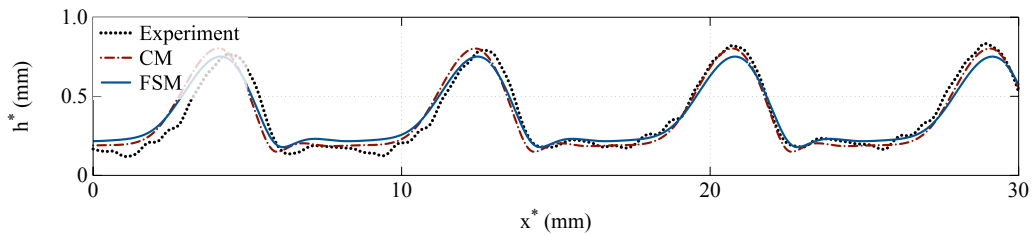}
      \caption{Film thickness comparison between experiment and theory, for fiber size $\R =0.215$ mm, flow rate $\Qm = 0.04$ g/s and nozzle inner diameter $ID = 3.0$ mm. }
   \label{comparison_profile}
\end{figure}

Figure \ref{fig:profile_thin_thick} shows the different morphologies that arise when using a thick fiber versus a thin fiber. In the former case, the droplet height \red{and width are}  close to the period length. In the latter case, the droplets appear more isolated. 
The profiles in Figure~\ref{comparison_profile} illustrate a direct comparison between the profiles predicted by both models, which are within the margin of error from the experimental film thickness. 

\begin{figure}
   \centering
   \begin{tabular} {ccc}
   \subfloat[$\R = 0.10$ mm, $\Qm = 0.04$ g/s]{  \includegraphics[width= 0.5\textwidth]{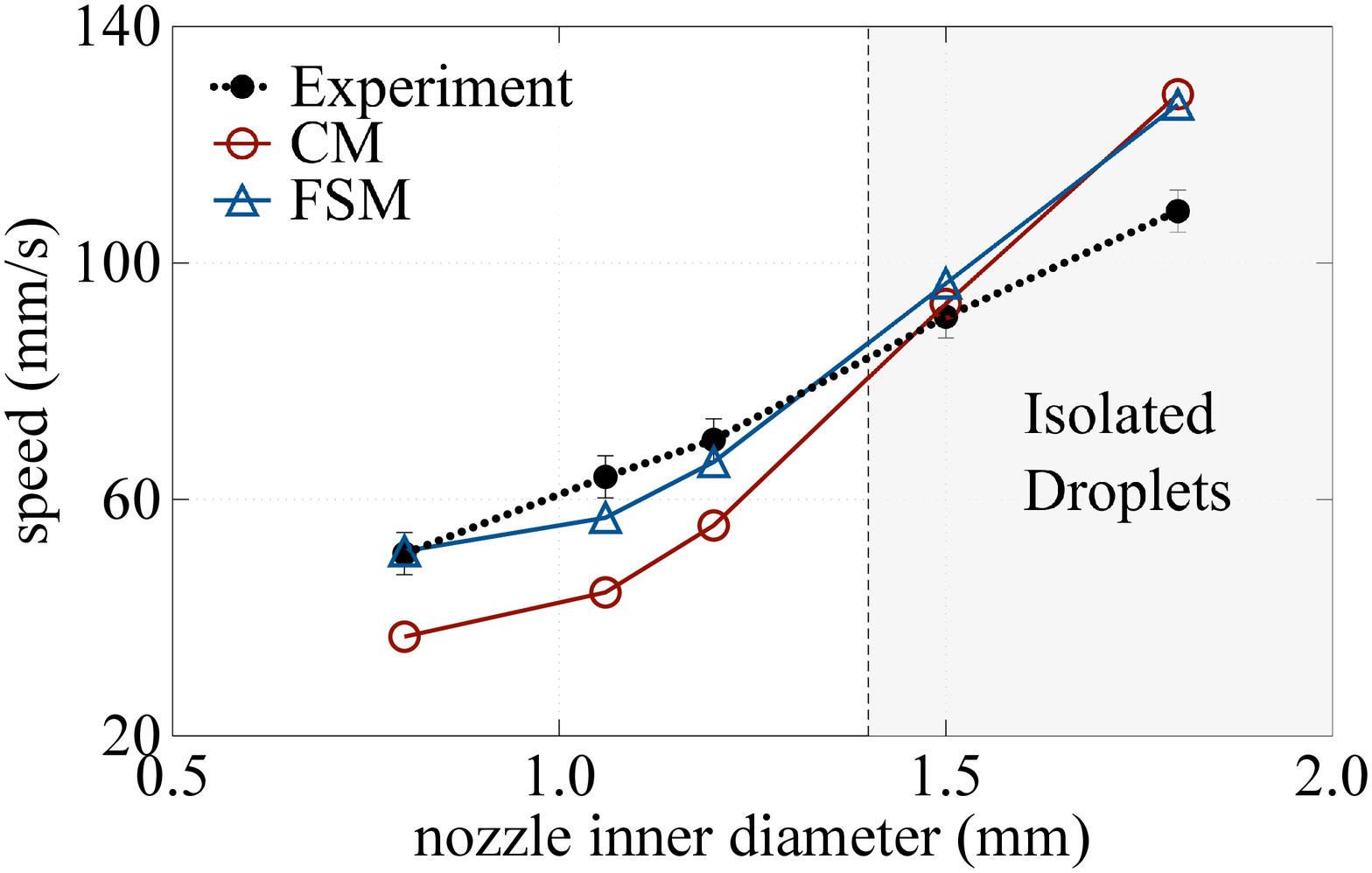} } &
     \subfloat[$\R = 0.215$ mm, $\Qm = 0.04$ g/s]{  \includegraphics[width= 0.5\textwidth]{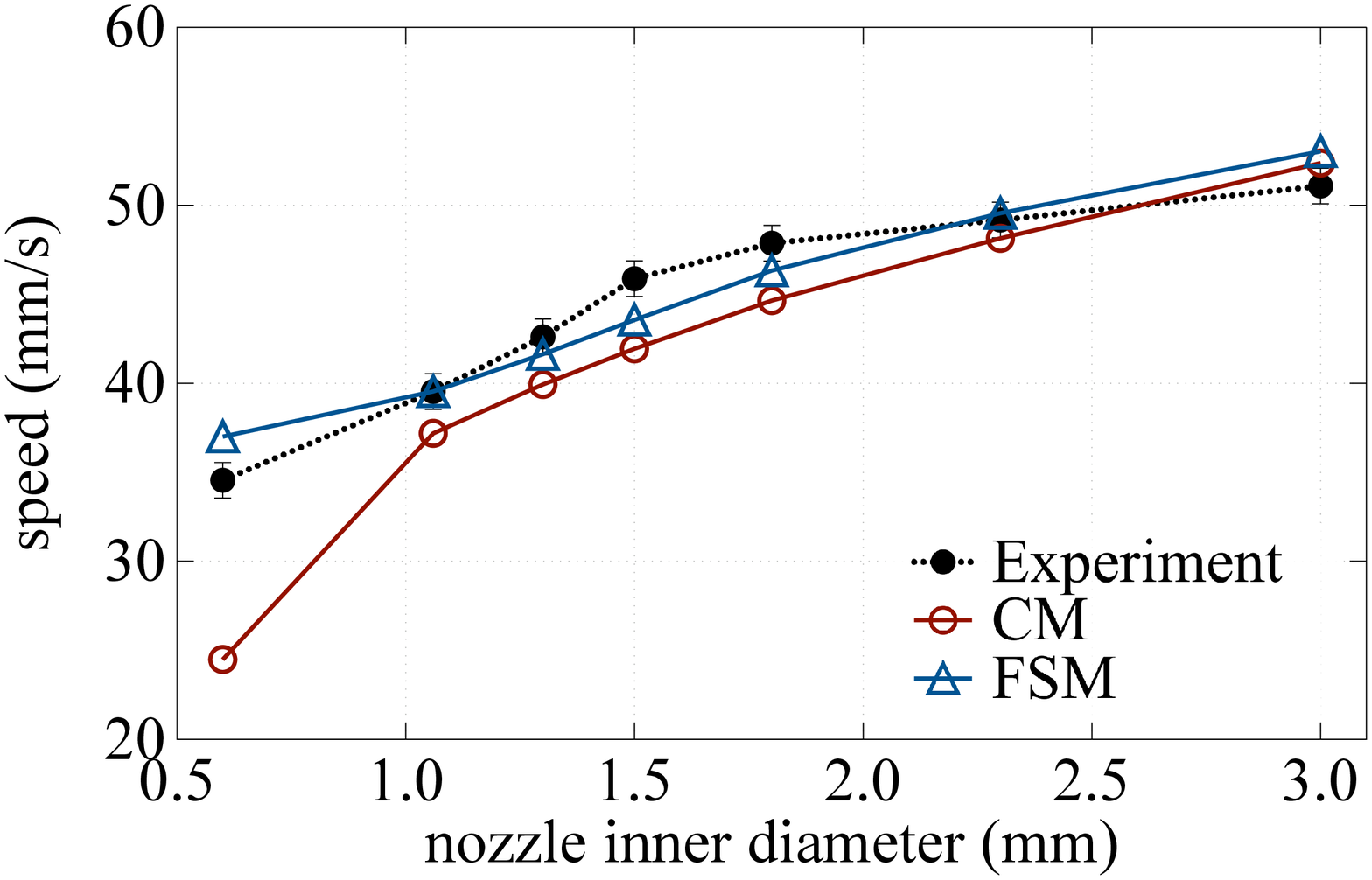} } &\\
    
    \subfloat[$\R = 0.10$ mm, $\Qm = 0.06$ g/s]{  \includegraphics[width= 0.5\textwidth]{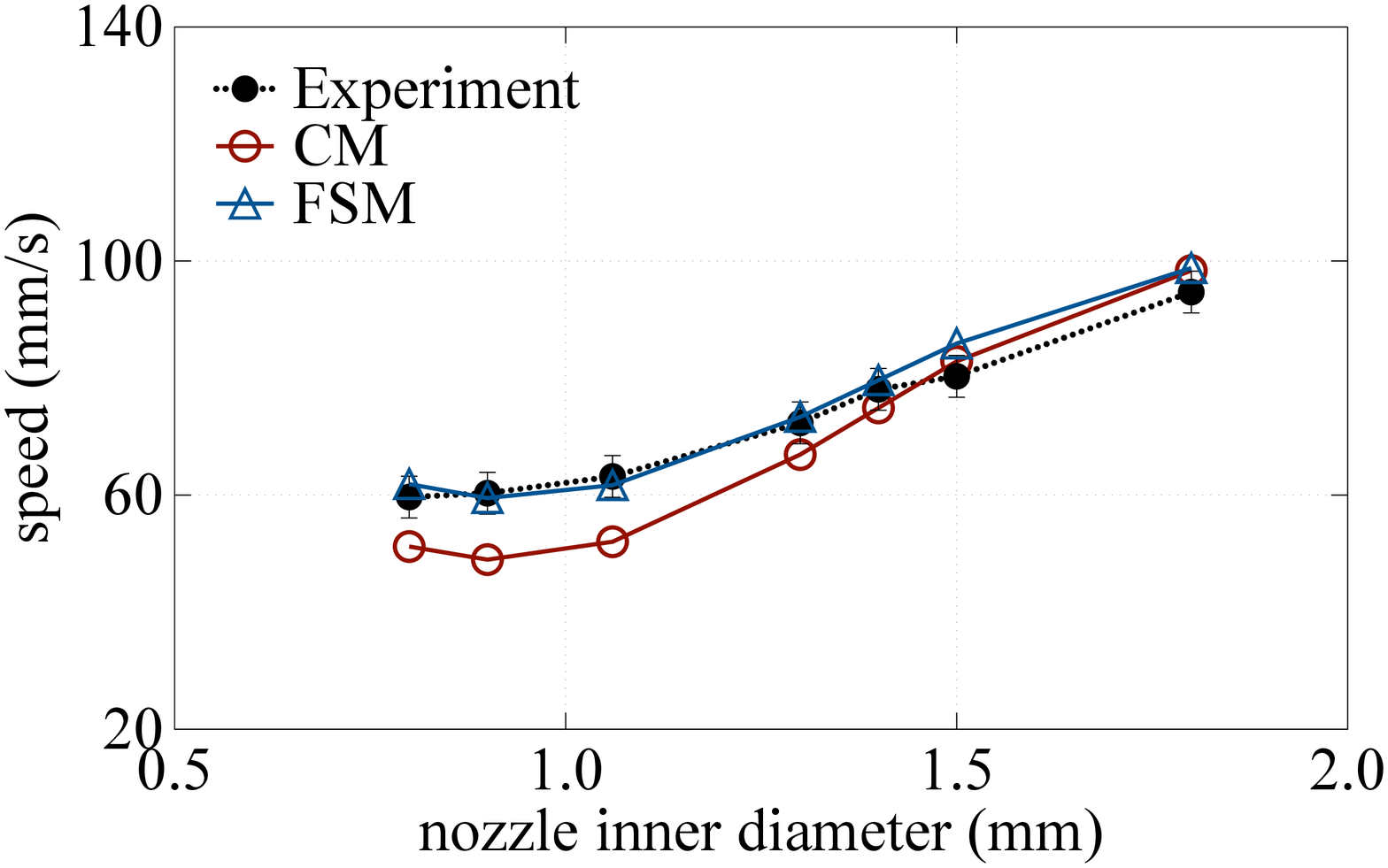} } &
   \subfloat[$\R = 0.215$ mm, $\Qm = 0.06$ g/s]{  \includegraphics[width= 0.5\textwidth]{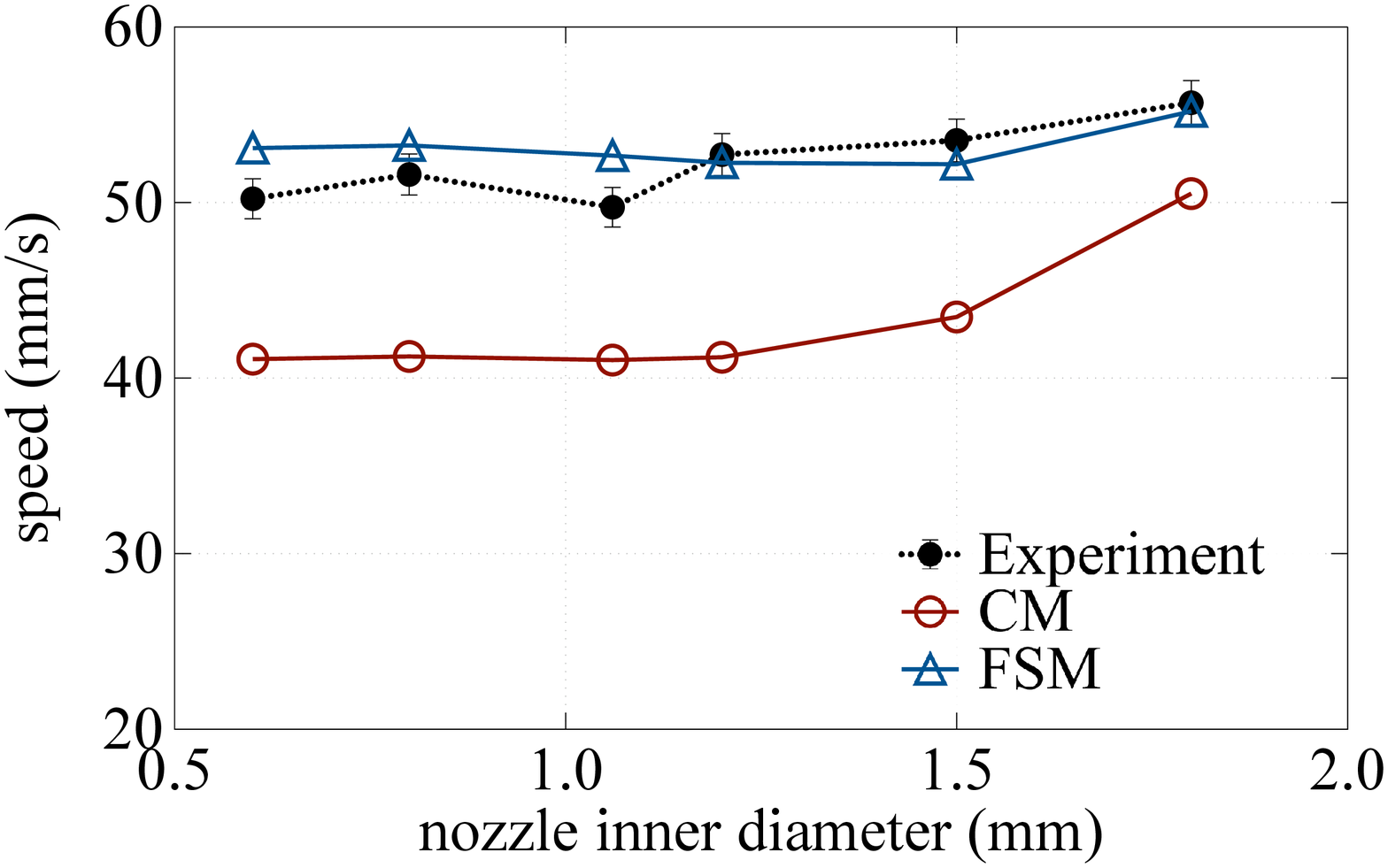} }   & \\
    
    \subfloat[$\R = 0.10$ mm, $\Qm = 0.08$ g/s]{ \includegraphics[width= 0.5\textwidth]{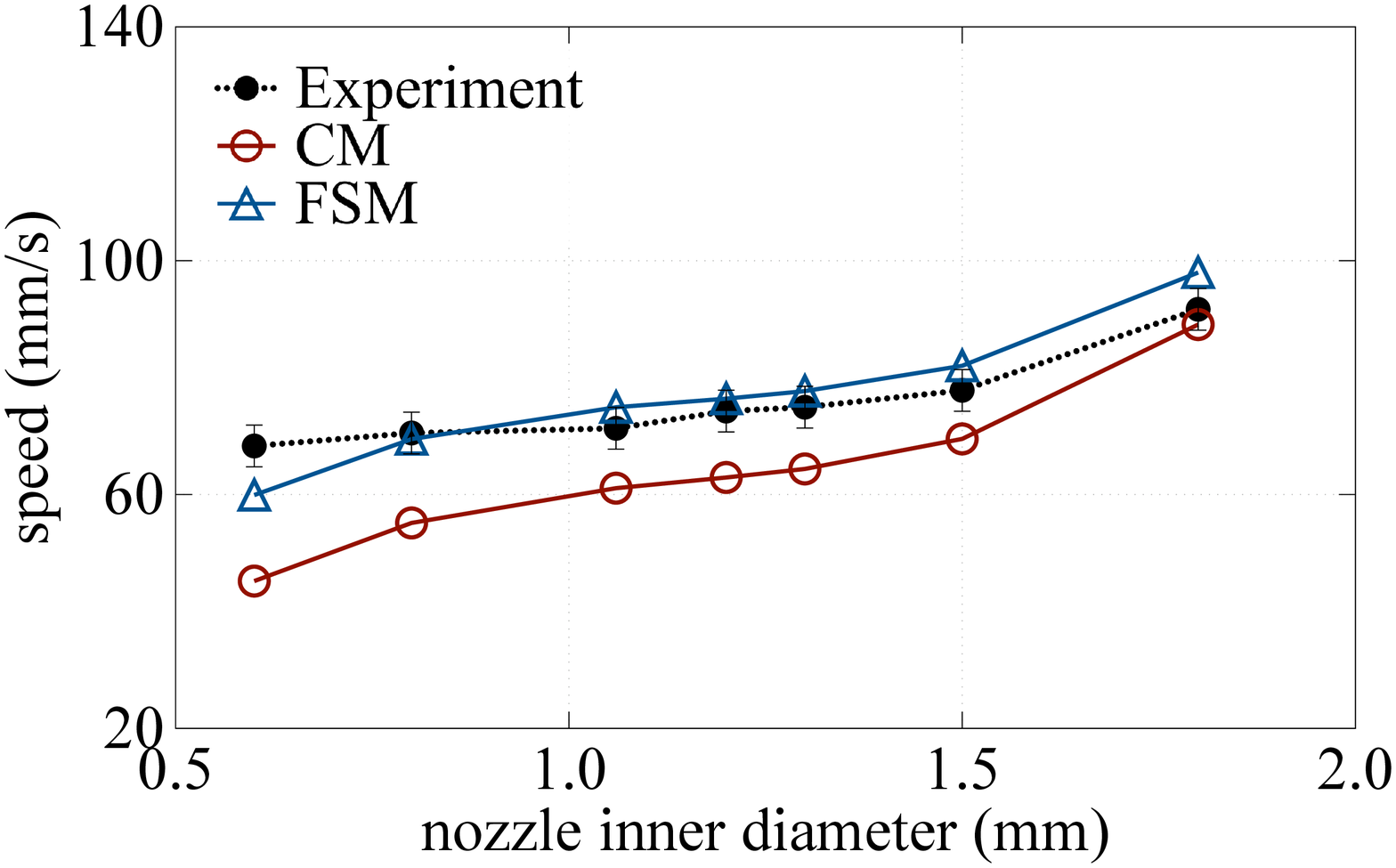} }
         & 
          \end{tabular}
   \caption{Average bead speed for fiber radius $\R=0.1$ mm (left) and $\R = 0.215$ mm (right), and flow rates $\Qm =$ 0.04, 0.06, and 0.08 g/s from top to bottom, compared to the proposed model (\GvdW) as blue diamonds and Craster \& Matar (CM) as red circles. The last two data points in (a) are in the isolated droplet regime. There is no plot for $\R =0.215$ mm, $\Qm = 0.08$ g/s, because in this case there is no TWS and the experiments lie in the convective regime. }
   \label{experiment_speeds}
\end{figure}

The predicted speeds, however, have noticeable differences between the two models. In Figure \ref{experiment_speeds}, we show plots of the observed and predicted speeds for varying nozzle sizes. The left panels indicate the thin fiber $ \R = 0.1$ mm whereas the right panels illustrate the thick fiber $\R = 0.215$ mm, with the flow rate $\Qm$ increasing from top to bottom. The \GvdW model agrees quite well with the experimental observations across all of the data. In contrast, the CM model underestimates the speed. We do not show the case with the fiber radius $\R = 0.215$ mm and the flow rate $\Qm = 0.08$ g/s because under those conditions the flow is in the convective regime and the bead speed cannot be uniquely defined. We note that for the \GvdW model, we choose $\epsilon_p^*$ to be $0.12$ mm for the fiber of radius $\R = 0.1$ mm, and $\epsilon_p^* = 0.215$ mm for the fiber of radius $\R = 0.215$ mm.

Note in Figure \ref{experiment_speeds}(a), the Rayleigh Plateau regime transitions to the isolated droplet regime, where the TWS is unstable and both models over-predict the speed, as expected. The following section discusses this transition in detail.

\subsection{Regime Transition} \label{sec:regime_transition}
 
\begin{figure}
\centering
{\includegraphics[width = 0.8\textwidth]{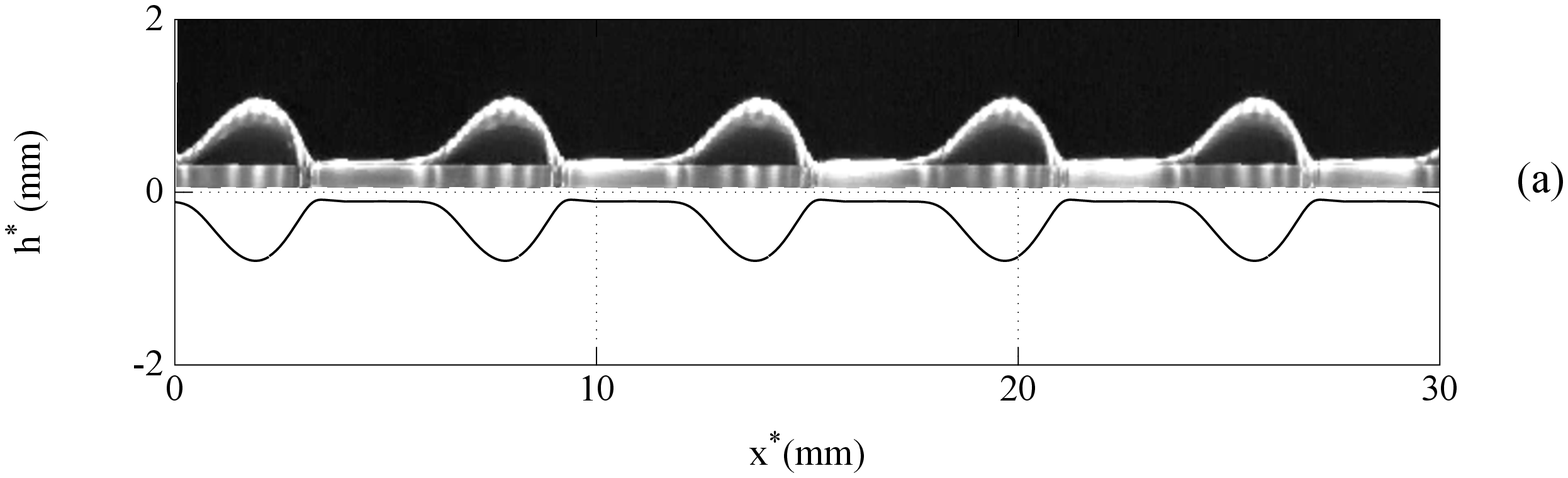}}\\
{\includegraphics[width = 0.8\textwidth]{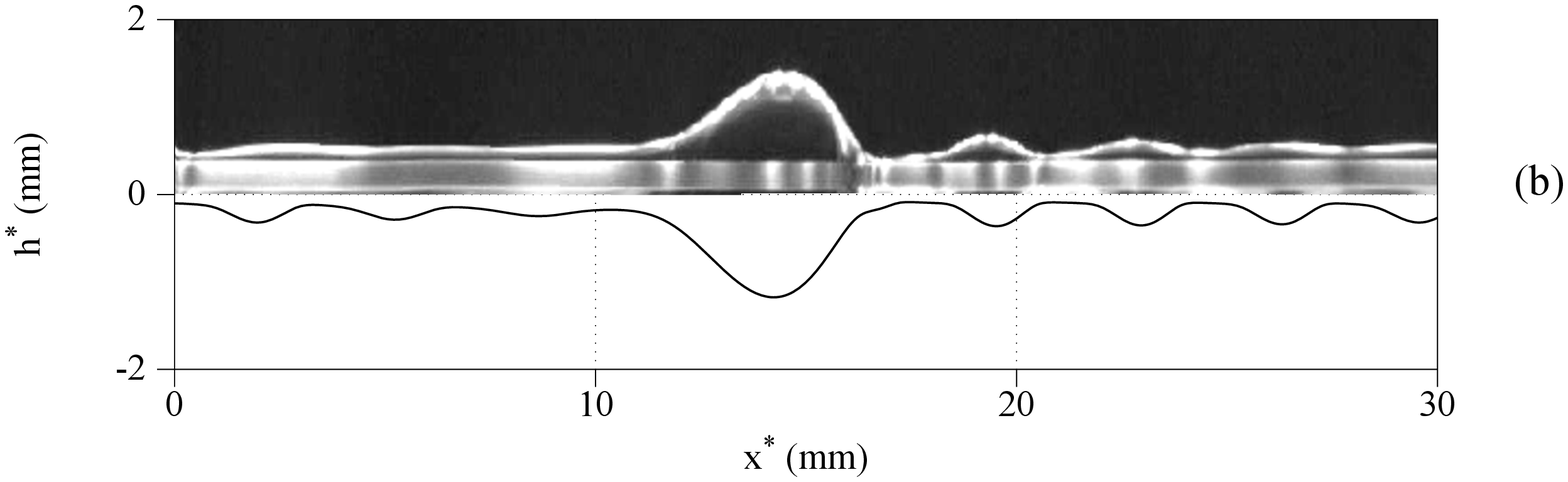}}
\caption{ Experiments (top black halves) with parameters $\R =0.1$ mm, $\Qm = 0.04 $  g/s against the \GvdW (bottom halves) showing (a) the Rayleigh-Plateau regime with nozzle $ID =$ 0.8 mm and (b) the isolated droplet regime with nozzle $ID =$ 1.8 mm.  }
\label{RP_and_IS_regime_profiles}
\end{figure}

 The emergence of instabilities in the thin liquid films between traveling droplets characterizes the transition from the Rayleigh-Plateau (RP) regime to the isolated droplet (IS) regime. Guided by the stability analysis from section \ref{sec:gvdw}, we can explore these instabilities and the departure from the RP regime as the nozzle diameter varies. The IS regime gives rise to time periodic dynamics that cannot be captured by the traveling wave solutions of \eqref{eq:travelingODE}. In this case, we need to solve the fully time-dependent model \eqref{eq:main}.
The numerical solution and the experimental observation from the IS regime are plotted in Figure \ref{RP_and_IS_regime_profiles} (b). In contrast, the traveling wave solution for an RP experiment ($ID = 0.8$ mm) is plotted in Figure\ref{RP_and_IS_regime_profiles} (a). \red{ Our dynamic simulation captures a difference in the advancing and receding lines of the droplets, even though this difference is more pronounced in the experiment.}

\begin{figure}
\centering
\subfloat[]{\includegraphics[width =0.28\textwidth]{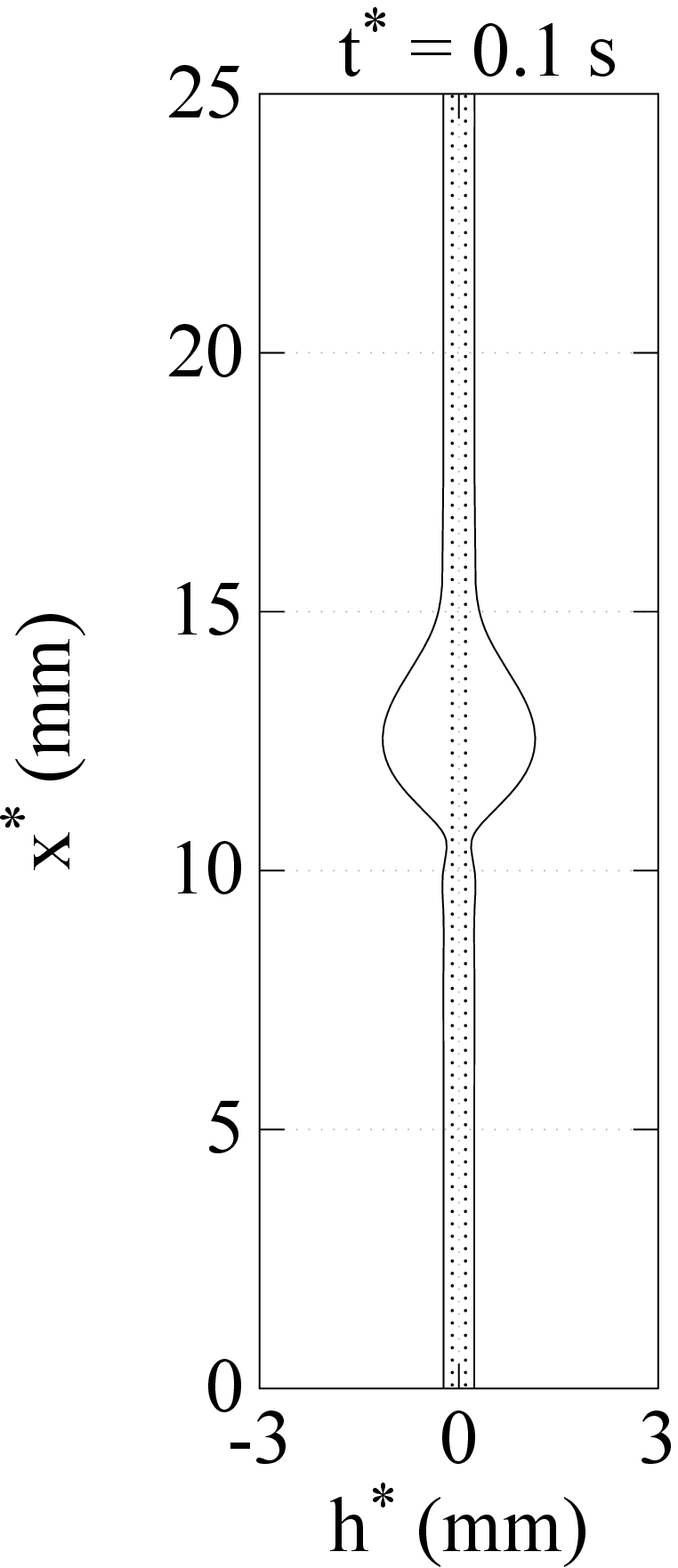}}~
\subfloat[]{\includegraphics[width =0.28\textwidth]{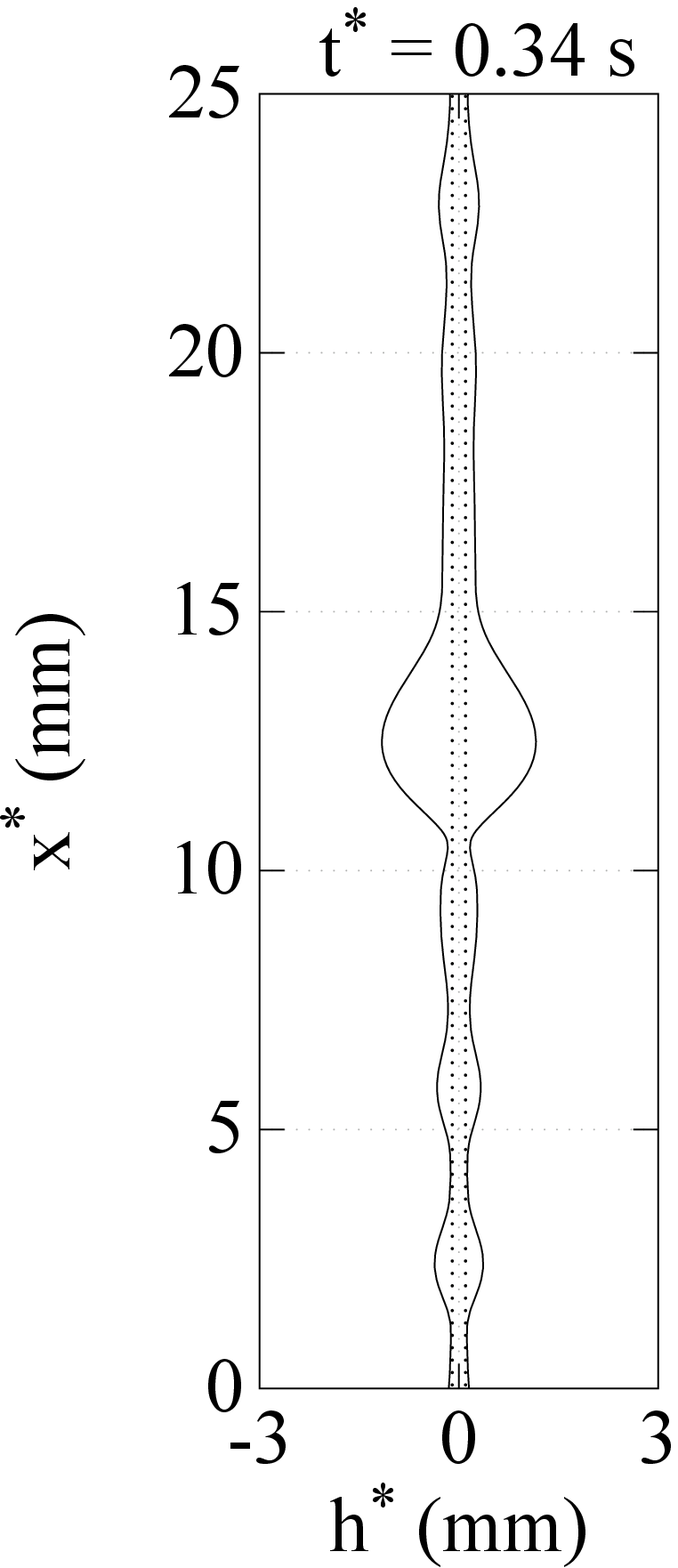}}~
\subfloat[]{\includegraphics[width =0.28\textwidth]{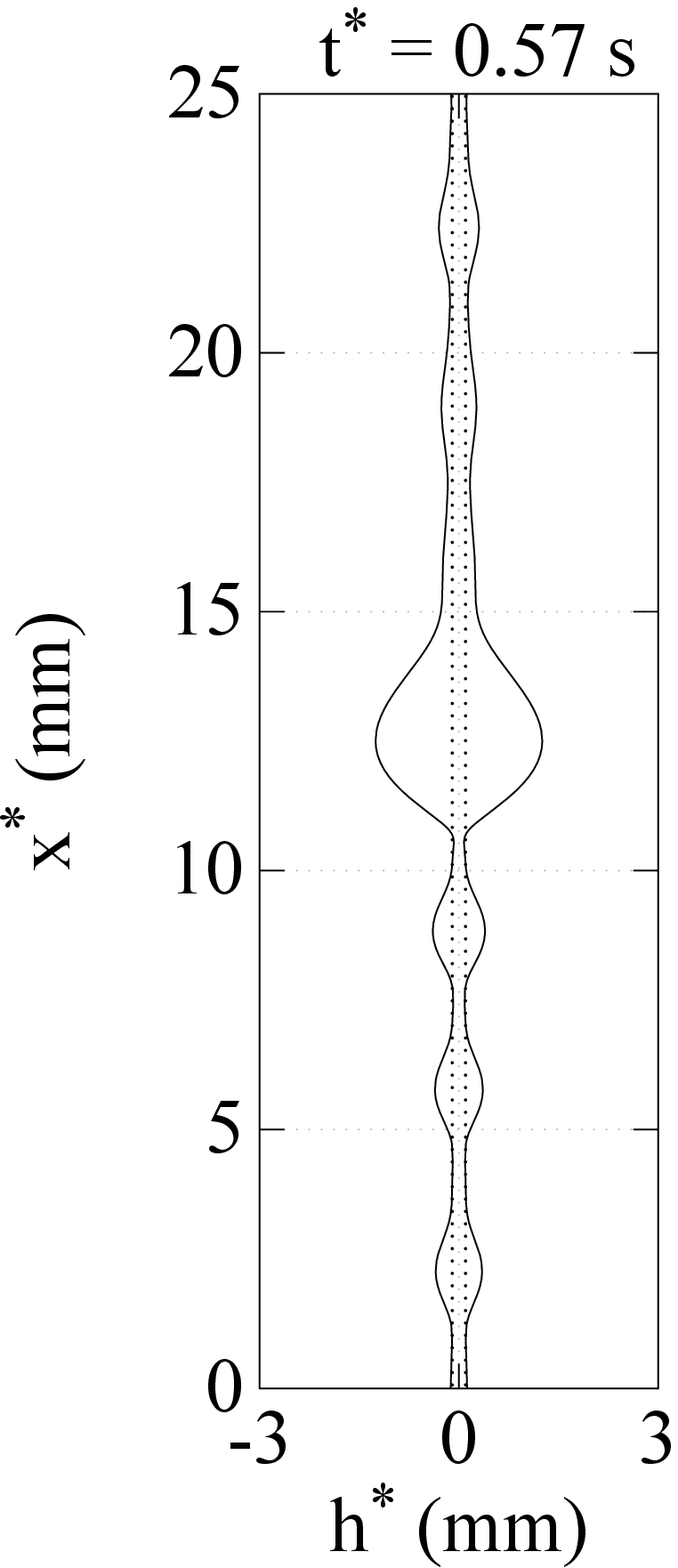}}
\caption{ Dynamic simulation showing the development of instabilities in the isolated droplet regime for parameters $\R = 0.1$ mm, $\Qm = 0.04$ g/s, and nozzle size $ID= 1.8 $ mm. }
\label{ISRegimeSimulation}
\end{figure}

The nonlinear PDE \eqref{eq:main} was solved numerically using a fully implicit second-order finite difference method. Figure~\ref{ISRegimeSimulation} shows the dynamics starting from a widely spaced traveling wave solution obtained from the ODE \eqref{eq:travelingODE} with a small perturbation, 
$
h_0(x) = H(x) + 0.001 \Psi(x).
$
This simulation corresponds to the IS regime experiment with the fiber radius $\R=0.1$ mm, flow rate $\Qm=0.04$ g/s and the nozzle diameter $ID=1.8$ mm. It demonstrates the effects of the unstable eigenmodes shown in Figure~\ref{stability_tws} (right). Such instabilities lead to interesting spatio-temporal pattern formation with small droplets appearing from the unstable long flat film between the large traveling beads. 

\begin{figure} 
\centering
\subfloat[\GvdW]{\includegraphics[width=2.6in]{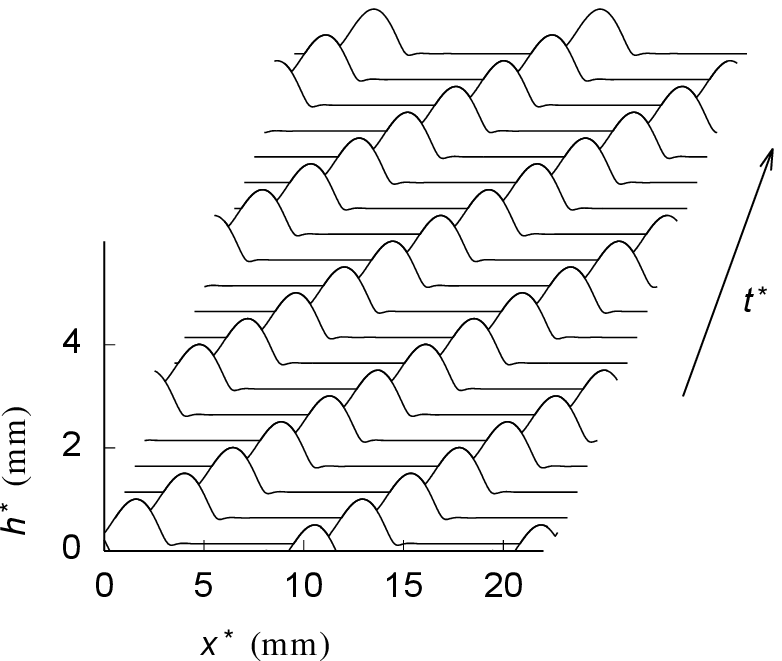}}
\subfloat[CM]{\includegraphics[width=2.6in]{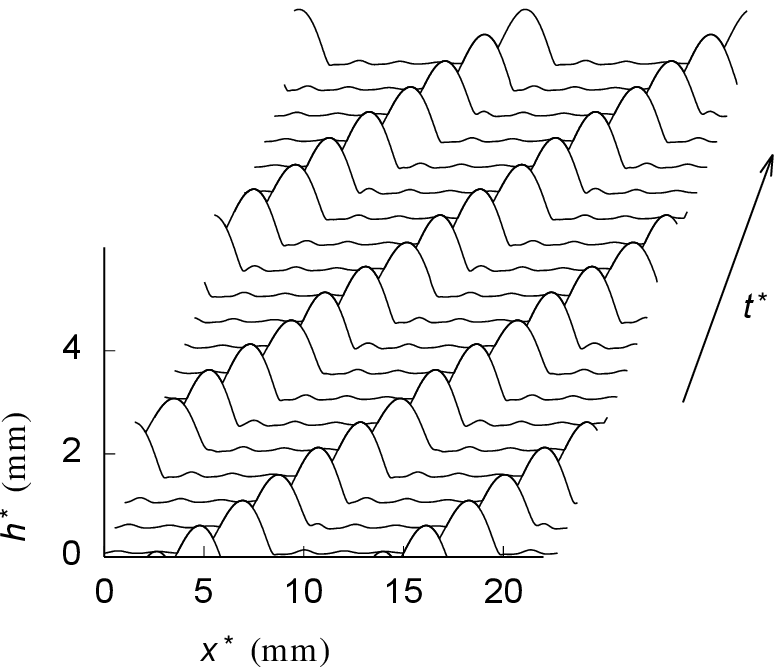}}
\caption{PDE simulations for (a) the \GvdW model \eqref{eq:main} with $\epsilon_p^* = 0.12$ mm and (b) the CM model starting from $h_0(x)=H(x)$ with identical $(M_0, L)$ and small perturbations, showing that the TWS to the \GvdW model is stable in time, while the one for the CM model is unstable and converges to a time-periodic solution in the long time. The $(M_0, L)$ values are obtained from the experiment with fiber radius $\R = 0.1$ mm, flow rate $\Qm = 0.04$ g/s and nozzle $ID = 1.2$ mm.}
\label{pde_comparison_CM_PCM}
\end{figure}

\begin{figure} 
\centering
\includegraphics[width=3in]{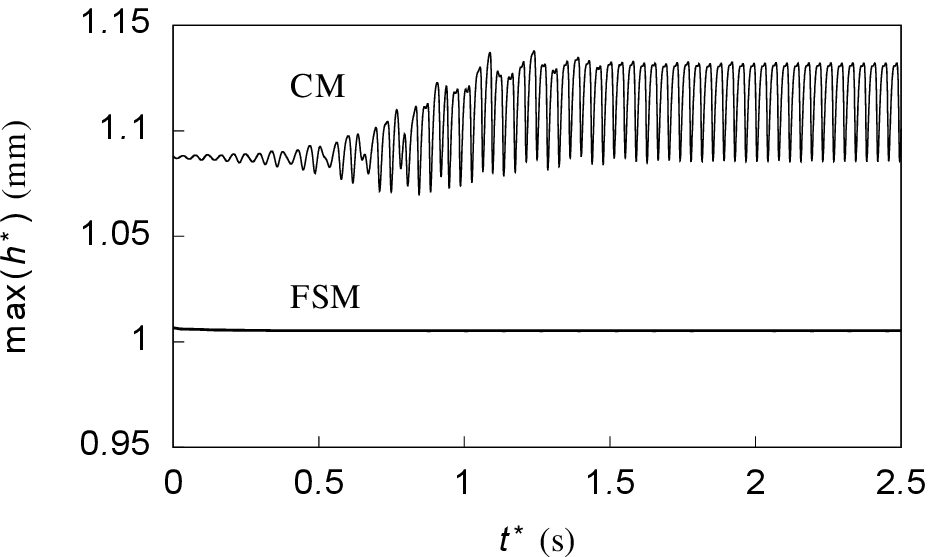}
 \caption{Plots of $\max(h^*)$ in time for the PDE simulations shown in Figure~\ref{pde_comparison_CM_PCM}.
}
\label{hmax_comparison_CM_PCM}
\end{figure}

 The film stabilization  model (\GvdW) correctly captures the bifurcation from the RP regime to the IS regime as the nozzle diameter increases. For small fiber size $\R= 0.1$ mm at $\Qm = 0.04$ g/s, experiments show that the regime transition occurs between nozzle inner diameters $1.2$ mm and $1.5$ mm (see Figure~\ref{experiment_speeds} (a)). However, the CM model predicts the transition between $0.8$ mm and $1.06$ mm, contradicting experimental observations. That is, based on stability analysis, the traveling wave solutions of the CM model are unstable for all nozzles bigger than $0.8$ mm. For example, for the case of the nozzle diameter $ID = 1.2$ mm, Figure~\ref{pde_comparison_CM_PCM} shows a comparison of dynamic simulation results of the CM and the \GvdW models. The traveling wave in the \GvdW model propagates steadily with a constant speed and profile (see the bottom curve in Figure~\ref{hmax_comparison_CM_PCM}). In contrast, the CM model simulation shows that the initially nearly-flat coating layer quickly evolves into small waves ahead of the major sliding bead. As the major bead interacts with the small waves downstream, the maximum height of the film thickness oscillates in time (see the top curve in Figure~\ref{hmax_comparison_CM_PCM}) since the main bead gains mass from these smaller waves. The dynamic solution eventually converges to a time-periodic solution that describes the case in the isolated droplet regime. A similar bifurcation occurs when keeping the nozzle diameter constant $ID= 1.8$ mm and changing the flow rate (not shown here). Again, just the \GvdW model correctly predicts regime transition between $\Qm = 0.050$ g/s and $\Qm = 0.055$  g/s, which agrees with experimental measurements. For larger fiber size $\R = 0.215 $ mm, both models produce a stable traveling wave for all nozzle sizes shown in Figure~\ref{experiment_speeds}, i.e. there is no isolated phenomena predicted.

\subsection{The effects of slip and curvature}
\label{sec:slipCurvature}

\begin{figure} 
\centering
\includegraphics[width=2.6in]{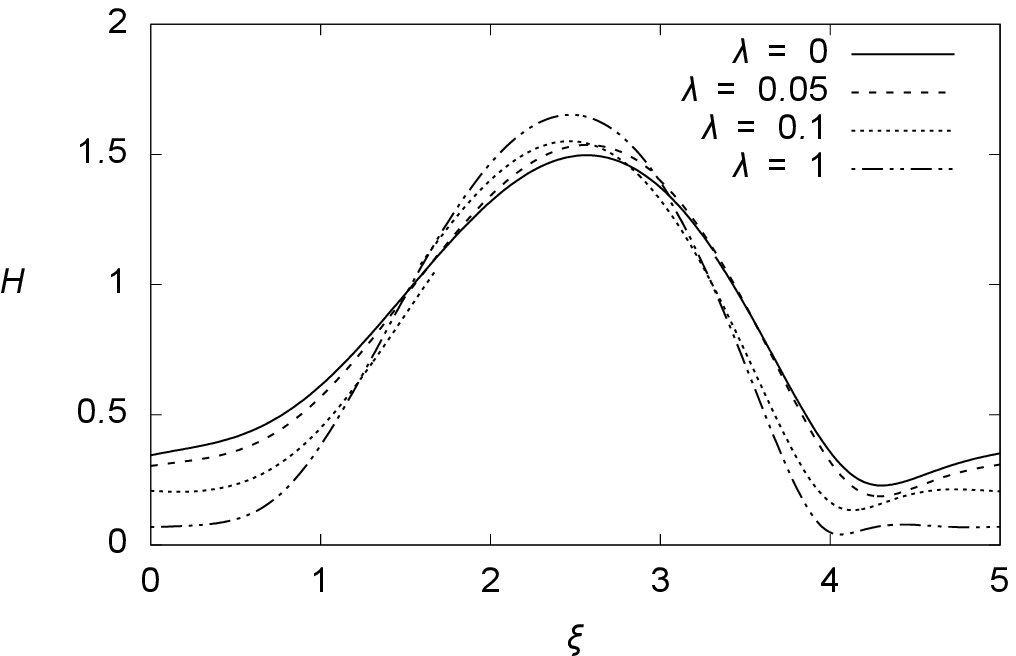}
\includegraphics[width=2.6in]{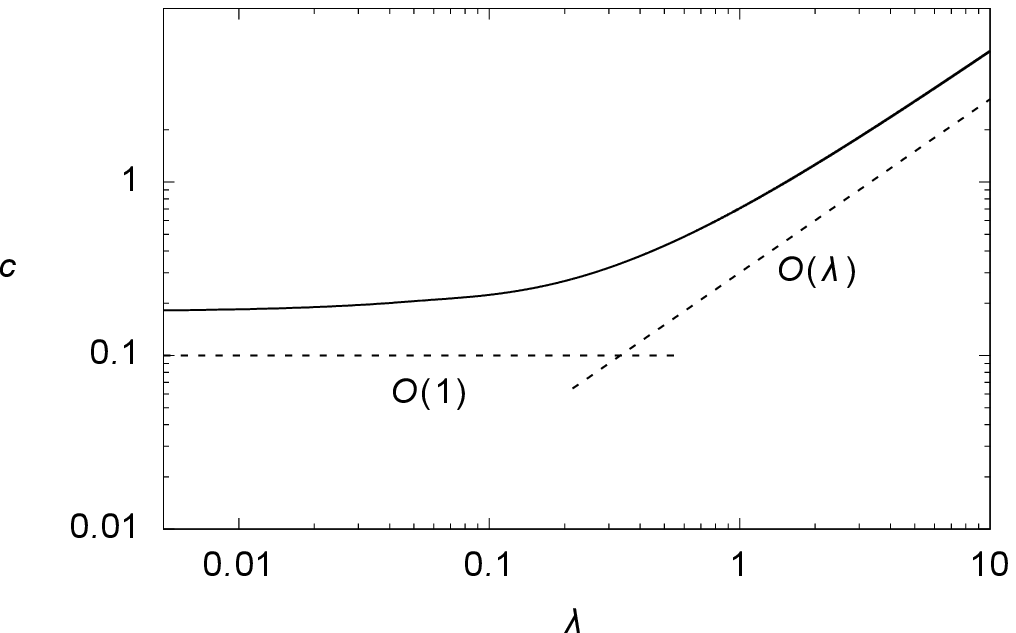}
   \caption{Traveling wave solutions of \eqref{eq:travelingODE} for the SCM model with a varying slip length $\lambda = 0, 0.05, 0.1, 1$ for $(L,M_0) = (4.94,8.27)$ showing that (left) larger slip yields larger droplet heights and thinner precursor layers (right) in the weak slip limit the moving speed of drops is $O(1)$, and the speed increases linearly with slip length in the large slip limit.}
   \label{slip_comparison}
\end{figure}

We next study how the slip length $\lambda$ affects the flow characteristics under the slip model (SCM).
It has been shown in \cite{halpern2017slip} that the presence of slip enhances capillary instability and promotes both droplet formation and the  speed of the falling drops.
Similar observations are also made in our study of traveling wave solutions for given $(L,M_0)$ in Figure~\ref{slip_comparison}.
We observe that compared with the no-slip case, 
the  slip cases ($\lambda \neq 0$) have taller and narrower droplets.
The dependence of the speed $c$ on the slip length $\lambda$ in Figure~\ref{slip_comparison}~(right) agrees with the observation in \cite{halpern2017slip}. 
The speed of traveling wave solutions is of order $O(1)$ in the weak slip limit and grows linearly with $\lambda$ in the strong slip limit. 
Since typical slip length $\lambda^*$ is $1 - 10~\mu$m for silicone oil, and typical Nusselt film thickness $\mathcal{H}$ is about $0.5$ mm in our experiments, our experimental conditions are expected to be in  the weak slip limit.

\begin{figure} 
\centering
\includegraphics[width=0.6\textwidth]{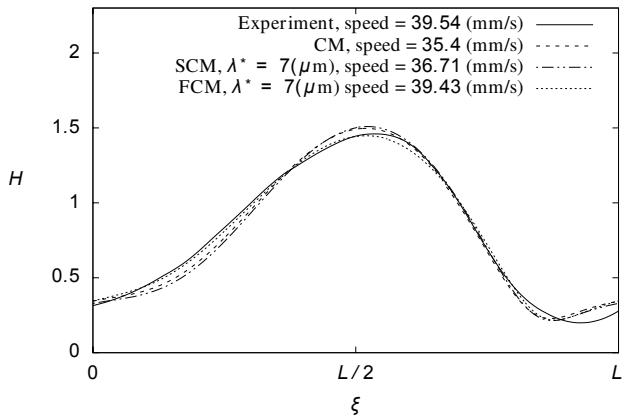}
\caption{A comparison of traveling wave solutions to \eqref{eq:travelingODE} from the original Craster \& Matar model (CM), the Slip Craster \& Matar model (SCM), the Full Curvature Model (FCM) and the experimental observation showing that while the solutions profiles from each models are close, the estimated bead velocity predicted by the FCM matches the best with the experimental result. The bead profile is obtained from the thick fiber experiment with $\R = 0.215$ mm, flow rate $\Qm$ $=0.04$ g/s, and the inner nozzle diameter ID = $1.06$ mm. The corresponding non-dimensional constraints are given by $(L, M_0) = (4.94,8.92)$.}
 \label{curvature_comparison}
\end{figure}

\begin{figure}
   \centering
  \includegraphics[width= 0.45\textwidth]{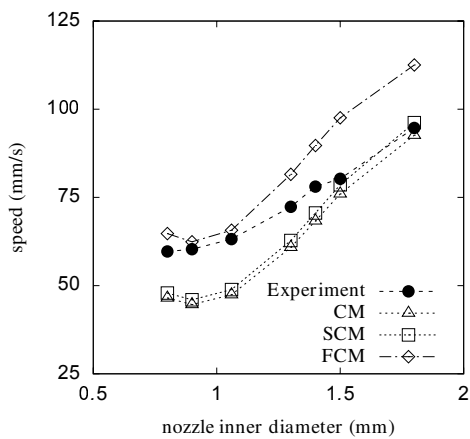} 
   \caption{Average bead speed obtained from experiments with $\R = 0.1$ mm and $\Qm  = 0.06$ g/s, compared to the speed predicted by the original CM model, the SCM model with $\lambda^*=3$ $\mu$m, and the FCM model with $\lambda^*=3$ $\mu$m.}
 \label{allmodels_speeds}
\end{figure}

Similar to the slip model, the inclusion of the fully nonlinear curvature term in the FCM model with $\mathcal{Z}(h)$ given by \eqref{z2} also increases the moving speed of sliding droplets. Figure~\ref{curvature_comparison} shows a comparison of the experimental bead profile against those obtained from the Craster \& Matar model (CM), the slip model (SCM) with the linear azimuthal term in \eqref{z1}, and the full curvature model (FCM) with the full azimuthal curvature term in \eqref{z2}. Under the period and mass constraint, whereas the different models all yield solution profiles similar to the experimental result, the predicted speed obtained from the ODE \eqref{eq:travelingODE} is increased when the slippage effects and full curvature term are included, and the FCM model with the slip length $\lambda^*=7$ $\mu$m provides the best agreement with the experiment.

While the slip and the full curvature do influence predicted wave propagation velocities, their corrections alone are not sufficient to improve agreement with the experimental data.  Figure \ref{allmodels_speeds} shows that for the thin fiber case  ($\R = 0.1$ mm), the CM and SCM models underestimate the bead velocities for all nozzle sizes. Even though the FCM model yields a good agreement with the experiment for small nozzles, it overestimates the speed for large nozzles. Similar trends are observed for  the thick fiber case  ($\R = 0.215$ mm) (not shown here). 
In these cases, the film stabilization  term becomes vital for correctly predicting speeds and the regime transition, as discussed in section \ref{sec:correction_model}.

\section{Conclusion}
\label{sec:conclusion}

We have performed a thorough study of viscous flow down vertical fibers, comparing a range of experimental results to different models of interest. 
We focus on understanding the Rayleigh-Plateau (RP) regime, where traveling wave solutions emerge, and its transition to the isolated droplet (IS) regime due to nozzle effects. We propose a full model that incorporates the stability of thin uniform layers by including a film stabilization  (\GvdW) term to the pressure. 
In this paper, experiments are compared to versions of a lubrication model, showing the influences of slip effects, nonlinear azimuthal curvature and the \GvdW term on predicted bead velocities and film profiles.
In addition, we perform a stability analysis of the traveling wave solutions(TWS) that confirms  the importance of the \GvdW term to properly capture regime transitions. 
The model equations lead to both closely spaced wavy solutions and widely spaced droplet solutions which are affected by different fiber sizes, flow rates, and nozzle geometry. The slip model (SCM) leads to an increased speed of propagation and promotes formation of droplets. The full curvature model (FCM) also increases the magnitude of the bead velocity, while the \GvdW model stabilizes the thin undisturbed layer between consecutive droplets which leads to a more accurate speed prediction. Our results show outstanding experimental agreement using the \GvdW's TWS in the RP regime. In the isolated droplet regime, dynamic simulations of widely spaced droplet solutions agrees very well with experiments.

For future work, we are interested in the nozzle effects on the global modes  (\cite{duprat2007absolute}) of the absolutely unstable flows in RP and IS regimes. Of particular interest would be the selection of spacing between moving beads with a given nozzle diameter. \cite{duprat2009spatial} investigated the spatial response of a film flowing down a fiber to inlet forcing using a system of coupled equations for the flow rate and the film thickness. We anticipate that a similar model can be used to study the nozzle effects. \red{Moreover, the nozzle effects in our experiments are also related to the dripping faucet problem (\cite{coullet2005hydrodynamical,dreyer1991route}) which studies the chaotic behavior of a dripping faucet. 
Motivated by these theories, in the future we would like to further study the connection between the fluid dynamics near the nozzle and its influence on the downstream flow transitions.
} 

\red{ In addition, our experimental observations with different fluids, not published here, show that the nonlinear dynamics can change quite drastically, and choosing the appropriate coating thickness $\epsilon_p^*$ is an interesting question for further study.
For $\epsilon_p^*$ of the same order of the fiber radius, the value of the stabilization parameter $A^*$ is between $10^{-11}$ and $10^{-12}$ Nm, which is much larger than typical values of Hamaker constant between $10^{-19}$ and $10^{-21}$Nm. This indicates that the film stabilization term in our model is stronger than typical van der Waals interactions, and the underlying physics still needs future investigation.
}

This work was supported by the Simons Foundation Math+X investigator award number 510776 and the National Science Foundation under grant CBET-1358034.

\appendix

\section{Nomenclature}\label{app:Nomenclature}
In Table~\ref{table:nomenclature}, we show the relevant nomenclature and the corresponding values given a sample flow rate for thin and thick fibers. The first portion of the table shows dimensional parameters and their units, while the bottom part of the table shows the selected non-dimensional scales.


\begin{table}
\centering
\begin{tabular} {lcccccc}
 Definition&& Symbol&& Thin fiber&& Thick fiber\\
 &&&&&&\\
 Radius of the fiber (mm)&& $\R$ && 0.100 && 0.215 \\
sample flow rate (g/s) && $\Qm$  && $0.080$ && $0.040$ \\
nozzle inner diameter (mm) && $ID$  && $0.6 - 1.8$ && $0.6 - 3.0$ \\
average maximum height (mm) && $h_m^*$  && $0.8 - 1.0$ && $0.6 - 0.8$ \\

 lengthscale in radial direction (mm)  && $\mathcal{H}$  && 0.589 && 0.494\\
 lengthscale in streamwise direction (mm) && $\mathcal{L}$ && 1.091 && 1.029\\
characteristic streamwise velocity (mm/s)  &&$\mathcal{U}$ && 67.89
 && 47.87 \\
  capillary length (mm) &&$l_c$  && 1.485  && 1.485  \\

&&&&&&\\
aspect ratio $\mathcal{H}/\R$  && $\alpha$ && 5.886 && 2.299  \\
slip length   && $\lambda$ && 0.012 && 0.014 \\
scaling parameter $(\mathcal{H}/\mathcal{L})^2$  &&$\eta$  && 0.291 && 0.231   \\
uniform layer thickness  &&$\epsilon_p$  && 0.204 &&  0.435 \\
stabilization parameter   &&$A$  && 0.014 && 0.068    \\
mass constraint &&$M_0$  && 14 -- 26   &&  6 -- 13 \\
period &&$L$  && 4 -- 9 && 3 -- 8 \\
speed of TWS &&$c$  && 0.10 -- 0.15&& 0.18 -- 0.27  \\
Reynolds number  && Re  && 1.481 && 0.985  \\
Weber number   && We  && 2.522 && 3.004 
 \end{tabular}
  \caption{Nomenclature and their sample values.}
  \label{table:nomenclature}
\end{table}

\section{Nozzle effects} \label{app:nozzle}

 \begin{figure} 
\includegraphics[width= 0.45 \textwidth]{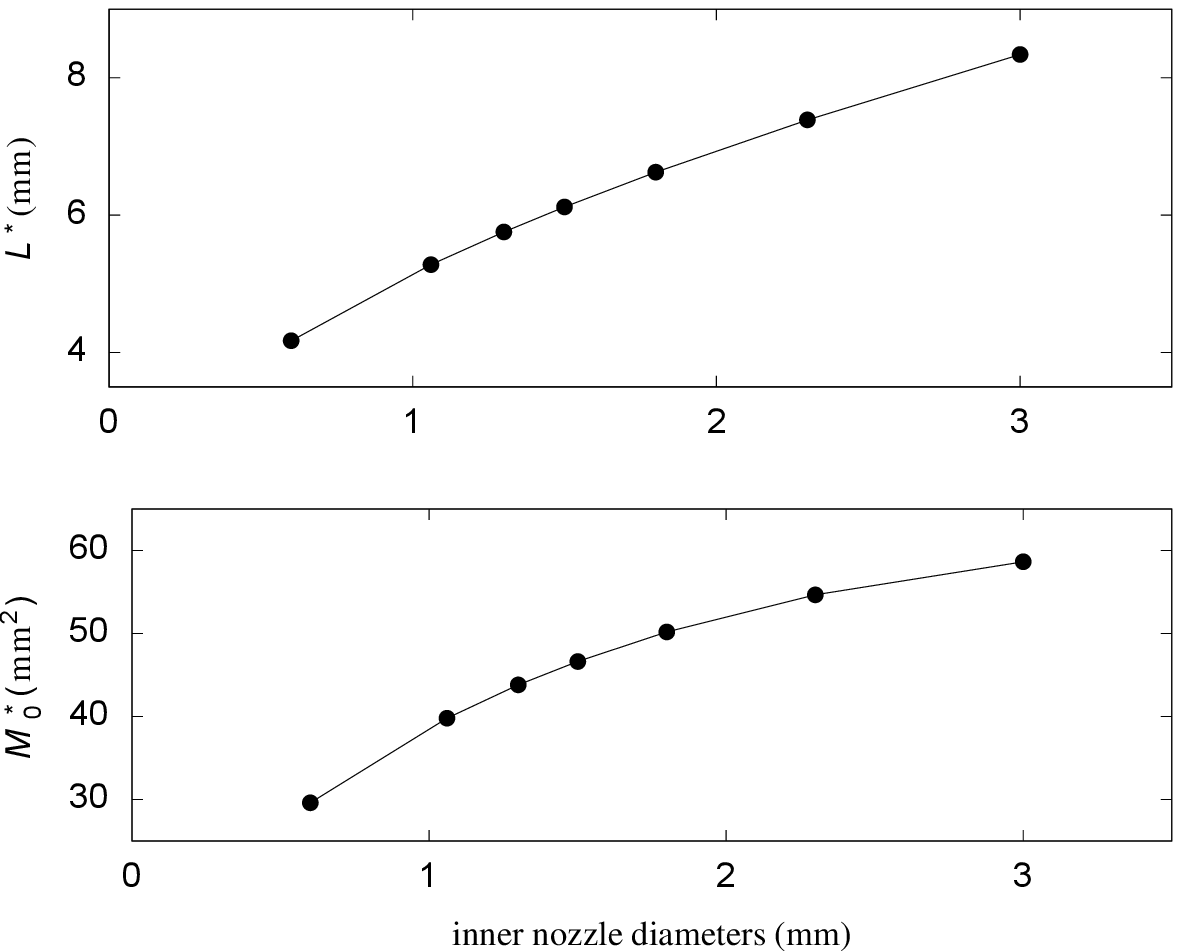}
\includegraphics[width= 0.55 \textwidth]{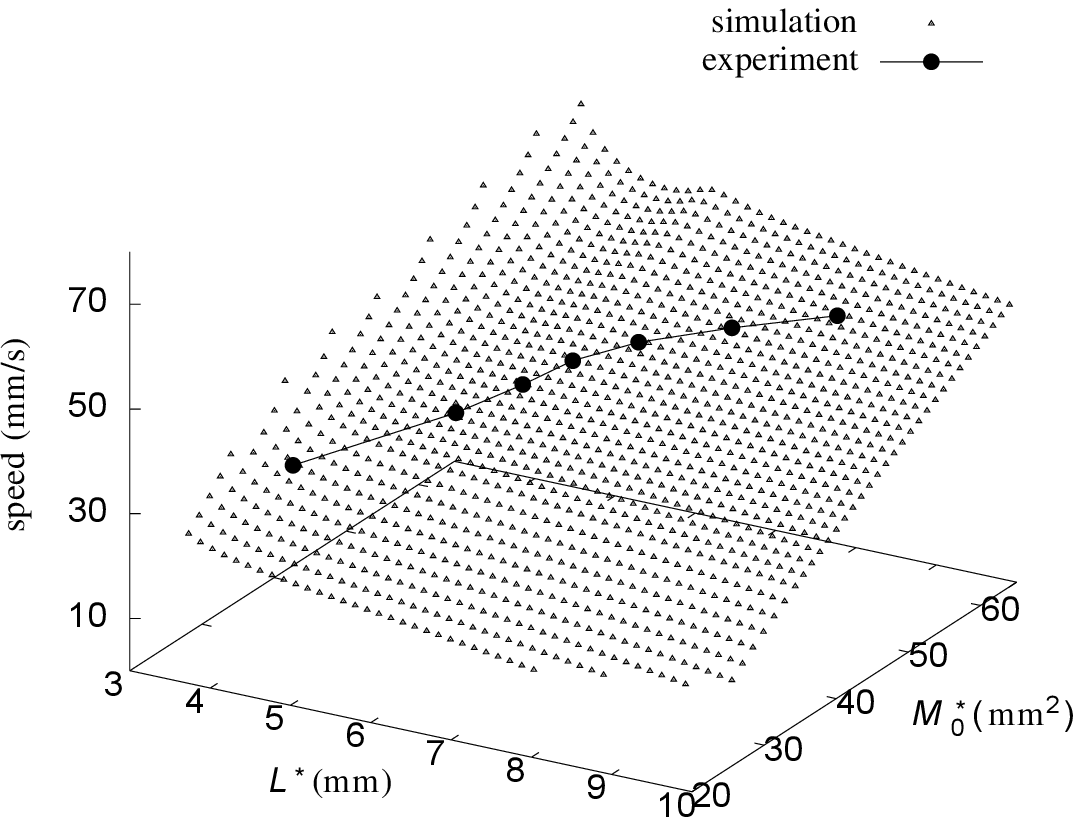}
\caption{ \red{(Left) The experimentally measured relations between the dimensional $L^*$, $M_0^*$, and the nozzle inner diameters for fiber radius $\R = 0.215$ mm at flow rate $\Qm = 0.04$ g/s; (Right) the predicted speed of traveling wave solutions to the film stabilization  model (\GvdW) in comparison with the corresponding experimental results (large dots) with varying $M_0^*$ and $L^*$}.}
   \label{3DplotLM0}
\end{figure}
Motivated by the different regimes shown in Figure~\ref{experiment_speeds}, we review the details of the dependence of bead speeds on its spacing and mass constraint, \red{and their relations to the nozzle diameters.} \red{ It has been empirically shown in \cite{sadeghpour2017effects} that when other experimental parameters are fixed, the nozzle diameter determines the spacing and mass constraint for the traveling beads. Typical ranges for the mass constraint and period with increasing nozzle inner diameter are shown in Table \ref{table:nomenclature}.  }

The plot in Figure~\ref{3DplotLM0} \red{(right)} shows the traveling wave speeds parametrized by a range of values for the dimensional period $L^*$ to examine the impact of large versus small bead spacings. We also use a range of values for $M_0^*$ to study the influence of bead sizes on its velocity. For each calculation with a given dimensional $(L^*,M_0^*)$, the numerically obtained traveling wave speed from the film stabilization  model is labeled by a small dot. 
For a fixed fiber radius at a given flow rate, the diameter of the nozzle feeding the fluid selects the $(L^*,M_0^*)$ pair. To illustrate this connection, Figure~\ref{3DplotLM0} \red{(right)} shows the experimental traveling bead velocities in large dots, for different nozzle sizes which correspond to specific $(L^*,M_0^*)$ pairs.
These results refer to experiments with fiber radius $\R= 0.215$ mm at flow rate $\Qm = 0.04$g/s, \red{and the measured relations between $L^*, M_0^*$, and the nozzle inner diameter are shown in Figure~\ref{3DplotLM0} (left)}. The nozzle diameter ranges from $0.6$ mm to $3.0$ mm. The period $L^*$ and the mass $M_0^*$ both increase as the nozzle diameter increases, and the predicted traveling wave speeds show good agreement with the experimental results.

\bibliographystyle{jfm}
\bibliography{FiberFilmsPaper}

\end{document}